\documentclass[twocolumn]{aastex6}
\usepackage{epsfig}
\usepackage{graphicx}
\usepackage{float}
\usepackage{amsmath}
\usepackage{color}
\usepackage{amssymb}
\usepackage{amsfonts}
\usepackage{units}
\usepackage{bm} 

\def\be{\begin{eqnarray}}
\def\ee{\end{eqnarray}}

\begin{document} 

\title{Fast radio bursts and their high-energy counterpart from magnetar magnetospheres}

\author{Yuan-Pei Yang\altaffilmark{1} and Bing Zhang\altaffilmark{2}} 

\affil{
$^1$ South-Western Institute for Astronomy Research, Yunnan University, Kunming, Yunnan, P.R.China; ypyang@ynu.edu.cn\\
$^2$ Department of Physics and Astronomy, University of Nevada, Las Vegas, NV 89154, USA; zhang@physics.unlv.edu
}

\begin{abstract} 

The recent discovery of a Galactic fast radio burst (FRB) occurring simultaneously with an X-ray burst (XRB) from the Galactic magnetar SGR J1935+2154 implies that at least some FRBs arise from magnetar activities. 
We propose that FRBs are triggered by crust fracturing of magnetars, with the burst event rate depending on the magnetic field strength in the crust. Since the crust fracturing rate is relatively higher in polar regions, FRBs are preferred to be triggered near the directions of multipolar magnetic poles.
Crust fracturing produces Alfv\'en waves, forming a charge starved region in the magnetosphere and leading to non-stationary pair plasma discharges. An FRB is produced by coherent plasma radiation due to nonuniform pair production across magnetic field lines. 
Meanwhile, the FRB-associated XRB is produced by the rapid relaxation of the external magnetic field lines. In this picture, the sharp-peak hard X-ray component in association with FRB 200428 is from a region between adjacent trapped fireballs, and its spectrum with a high cutoff energy is attributed to resonant Compton scattering. 
The persistent X-ray emission is from a hot spot heated by the magnetospheric activities, and its temperature evolution is dominated by magnetar surface cooling. Within this picture, magnetars with stronger fields tend to produce brighter and more frequent repeated bursts.

\end{abstract} 


\section{Introduction} 

Fast radio bursts (FRBs) are millisecond-duration radio transients with extremely high brightness temperatures \citep{Lorimer07,Petroff19,Cordes19,Zhang20}. Hundreds of FRB sources have been reported, dozens of which showed the repeating behavior (e.g., \citet{CHIME21}, and see the catalog of published FRB (http://frbcat.org, \citealt{Petroff16}) and Transient Name Server system for newly reported FRBs (https://www.wis-tns.org, \citealt{Petroff20})). Recently, a Galactic FRB with fluence of $1.5~{\rm MJy-ms}$ was detected to be associated with an X-ray burst (XRB) from the Galactic magnetar SGR J1935+2154 \citep{Bochenek20,CHIME20,Li20,Mereghetti20,Ridnaia20,Tavani20}, which suggests that at least some FRBs originate from magnetars \citep{Lyubarsky14,Katz16,Murase16,Beloborodov17,Kumar17,Metzger17,Yang18,Wadiasingh19,Lu20,Margalit20,Wadiasingh20,Wang20,Wang20b,Yang20,Ioka20,Zhang20,Xiao21,Lyubarsky21,Yu21}.

During the active phase of SGR J1935+2154 on 2020 April 28, over 200 XRBs were detected, and a burst storm with a rate of $>0.2~{\rm s^{-1}}$ appeared during the first 1120 seconds \citep{Younes20}. The $T_{90}$ duration distribution of the bursts peaks at $840~{\rm ms}$, and their spectra could be fit with a blackbody spectrum with an average temperature of $1.7~{\rm keV}$ and an emitting area of $53~{\rm km^2}$. Besides XRBs, the source also shows  persistent emission with a double-peaked pulse profile, with a period defined by the magnetar spin period. It is interesting that the arrival times of the two peaks of FRB 200428 aligned in phase with the brightest peaks of the associated XRB \citep{Li20} and with the peak of the pulse profile \citep{Younes20}. The flux and blackbody temperature of the X-ray persistent emission decreased rapidly in the early stages of the outburst with a duration of $\sim(0.1-1)~{\rm day}$, while the size of the emitting area remained unchanged \citep{Younes20}.

Although a large number of XRBs were detected on 2020 April 28, deep monitoring by Five-hundred-meter Aperture Spherical radio Telescope (FAST) suggested that the associations between FRB and X-ray burst are very rare \citep{Lin20}. Three possibilities might be considered to explain the association with low possibility \citep{Lin20}: 1. the radiation beaming angle of FRBs is much narrower than that of XRBs, so that there might exist off-beam ``slow'' radio bursts \citep{Zhang21}; 2. the peak frequency of FRBs has a wide distribution than the band of radio telescopes; 3. and the associations between FRBs and XRBs are intrinsically rare due to very special physical conditions \citep[e.g.][]{Li20,Younes20,Yang21}. 

For extragalactic FRBs, there is no confirmed multiwavelength counterpart so far. If an XRB is associated with an extragalactic/cosmological FRB at a distance $\sim 100~{\rm Mpc}$, the XRB flux would be much smaller than that of the XRB associated with FRB 200428, which is almost undetectable with the available high-energy detectors. Thus, identifying the magnetar origin of extragalactic FRBs via multi-wavelength observations would be extremely difficult. 
On the other hand, some extragalactic FRBs show special properties that are not similar to the properties of Galactic magnetars. For example, active repeating source FRB 180916B showed a period of $\sim16~{\rm day}$ with a frequency-dependent active window of $\sim 5$ days \citep{Marcote20,CHIME20b,Pastor-Marazuela20}. The first repeater FRB 121102 has mysterious persistent radio emission, a large, rapidly evolving rotation measure \citep{Spitler16,Chatterjee17,Michilli18} and a possible 160-day period \citep{Rajwade20}. Most magnetars in the Milky Way galaxy have a typical age of $\sim(1-10)~{\rm kyr}$. The extragalactic FRB sources may originate from magnetars with a wider age distribution at cosmological distances. In particular, those active repeaters may be related to magnetars with younger ages and/or stronger magnetic fields.

\begin{figure*}[]
    \centering
	\includegraphics[width = 0.9\linewidth , trim = 200 270 200 270, clip]{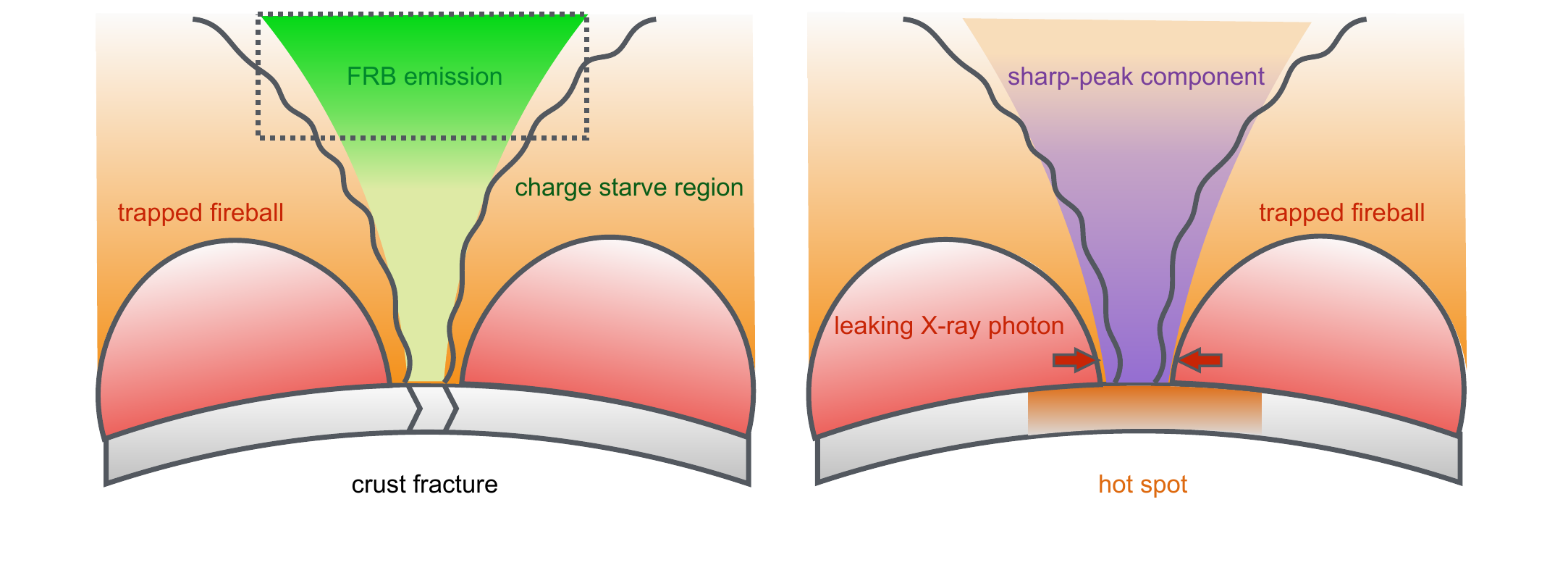}
    \caption{Schematic configuration. (a) FRB triggered by the crust fracture: a charge starved region is produced by Alfv\'en wave triggered by the crust fracture, and further generates FRB via coherent plasma radiation. XRB is produced by rapid relaxation of external magnetic field, and pair plasma generate and form trapped fireballs.  (b) sharp-peak X-ray component just after FRB: At the bottom of trapped fireballs, E-mode X-ray photons escape to the region between adjacent trapped fireballs, and are constrained in a beaming angle due to the large optical depth of the trapped fireballs around it, which corresponds to a sharp-peak component of XRB. Meanwhile, a part of XRB energy inject to magnetar surface. The magnetar surface is heated and generate persistent X-ray emission during a relatively long time. The dimensions in this figure are distorted for clarity.}\label{fig1}
\end{figure*}

The high-energy emission from a magnetar has been suggested to be due to magnetospheric activities \citep[e.g.,][]{Thompson95,Thompson01,Beloborodov07}. When a magnetar magnetosphere is triggered by crustal deformation, the magnetic energy in the crust would be released and be converted to particle energy and to radiation.
In this work, we conjecture that FRBs are triggered by crust fracturing of magnetars (see a schematic physical picture in Figure \ref{fig1}, see also \citet{Wadiasingh19} and \citet{Lu20}).
In the crust, the geometric configuration of magnetic fields evolves due to Hall drift. When the magnetic stress exceeds the shear modulus, the crust would crack. Since the crust fracturing rate is relatively larger in the polar regions with stronger magnetic fields, FRBs are preferably triggered near the directions of multipolar magnetic poles at the surface.
Crust fracturing produces Alfv\'en waves, and a charge starved region is generated in the magnetosphere \citep{Kumar20,Lu20}, leading to non-stationary pair plasma discharges. An FRB is produced by coherent plasma radiation due to nonuniform pair production across magnetic field lines \citep{Philippov20}. 
The FRB-associated XRB is produced by rapid relaxation of the external magnetic field.
In the polar region between adjacent trapped fireballs, a sharp-peak component of XRB emission is produced. Its spectrum is modified by resonant Compton scattering, and the radiation beam is constrained by the trapped fireballs surrounding it. After the FRB and XRB, a persistent X-ray emission component would be generated by a hot spot that is heated by the radiation near the magnetar surface.  
The paper is organized as follows. We first discuss the physical conditions and the event rate of crust fracturing in Section \ref{fracture}.
We propose that FRBs are produced by coherent plasma radiation in Section \ref{coherent}. 
The properties of FRB-associated XRB and persistent X-ray emission are discussed in Section \ref{Xray}. 
The results are discussed and summarized in Section \ref{summary}. The convention $Q_x = Q/10^x$ is adopted in cgs units, unless otherwise specified. 

\section{FRBs from crust fracturing and their burst rate}\label{fracture}

The association between FRB 200428 and an XRB from SGR J1935+2154 implies that FRBs might be triggered by an event that triggered the XRB. We attribute this trigger to crust cracking of the magnetar. According to this picture, FRBs with a wide range of luminosities are caused by a wide range of crust fracturing scales. 

The typical FRB event rate density is $\sim10^4~{\rm Gpc^{-3}yr^{-1}}$ with isotropic  luminosity $L>10^{42}~{\rm erg~s^{-1}}$, and the power law index of the cumulative luminosity function is $\alpha\sim-0.8$ for $N(>L)\propto L^\alpha$ \citep{Luo20}. This luminosity function seems to extend to as low as   $L\sim10^{38}~{\rm erg~s^{-1}}$, the isotropic luminosity of FRB 200428 associated with the XRB from SGR J1935+2154 \citep{Lu20}.
A magnetar may also emit radio bursts in an even wider luminosity/energy range. 
Indeed, in addition to FRB 200428, SGR J1935+2154 also emitted some radio bursts whose fluxes were several orders of magnitude 
lower than FRB 200428 \citep{Zhang20b,Kirsten20,Good20,Zhu20}. 

The evolution of magnetic fields in the crust triggers starquakes, further initiates magnetosphere activities, and generates high-energy emission (e.g., \citet{Thompson01,Beloborodov07}, see \citet{Wang18} for an FRB starquake model and \citet{Dehman20} for a simulation of crustal failures in young magnetars, although \citet{Levin12} argued that such fractures are unlikely to happen in magnetars.)\footnote{Besides internal magnetic field triggers, accretion processes in a binary system might also trigger crustal collapse and emit FRBs, which was recently considered in the scenarios invoking ultra-luminous X-ray sources \citep{Sridhar21}, Be/X-ray binary systems \citep{Li21c} and strange stars \citep{Geng21}.}. 
The magnetic energy inside the magnetar may be estimated as

\be
E_B\sim\frac{B^2}{8\pi}\left(\frac{4\pi}{3}R^3\right)\simeq1.7\times10^{47}~{\rm erg}~B_{15}^2,
\ee
where $R$ is the magnetar radius.
Assuming that the FRB emission makes use of the dissipated magnetic energy with an efficiency $\eta$, the intrinsic energy of one FRB should satisfy  $E_{\rm FRB} \eta^{-1} <  E_B$. For FRB 200428 from SGR J1935+2154, the radio-to-X-ray flux ratio is about $10^{-5}$, suggesting that the efficiency factor should be $\eta\lesssim10^{-5}$ \citep[e.g.,][]{Margalit20,Lu20,Zhang20}. Considering that the isotropic energy of FRB 200428 is $\sim 10^{35}$ erg, the magnetic energy of the magnetar is more than adequate to power this event, even if the efficiency is much lower, e.g. $\eta\lesssim 10^{-5}$.

For repeating FRB sources, one should address how many radio bursts could be produced within the energy budget of a magnetar. The first repeating source, FRB 121102, has been bursting nearly continuously for over 8 years. Its activity level significantly exceeds known Galactic magnetars. The FRB burst event rate is about (\citet{Wang19,Law17,Gourdji19}, also see Figure 3 of \citet{Margalit20})

\be
R(>E)\simeq R_0\left(\frac{E}{E_0}\right)^{-\alpha},
\ee
where $\alpha\sim 0.8$, $R_0\sim (10^4-10^5)~{\rm burst~yr^{-1}}$ is the average burst event rate with isotropic energy greater than $E_0\sim (10^{37}-10^{38})~{\rm erg}$ for FRB 121102 \citep[e.g.,][]{Li21b}. In general, FRBs may be beamed with an average solid angle $\Delta\Omega$. If the FRBs are on average emitted isotropically, the intrinsic event rate of FRBs would be $(4\pi/\Delta\Omega)R(>E)$, and the intrinsic energy of each burst would be $(\Delta\Omega/4\pi)E$. The two beaming factors are canceled out \citep[e.g.][]{Zhang20c}. Thus, the total released energy for a repeating source during the age $\tau$ would be 

\be
E_{\rm tot}(\tau)&\simeq&\tau\int_{E_0}^{E_{\max}} dE E\left|\frac{dR}{dE}\right|\simeq\frac{\alpha}{1-\alpha}\left(\frac{E_{\max}}{E_0}\right)^{1-\alpha}E_0 R_0\tau\nonumber\\
&=&1.0\times10^{47}~{\rm erg}\left(\frac{E_{\max}}{10^{40}~{\rm erg}}\right)^{0.2}\left(\frac{E_0}{10^{38}~{\rm erg}}\right)^{0.8}\nonumber\\
&\times&\left(\frac{R_0}{10^4~{\rm yr^{-1}}}\right)\left(\frac{\tau}{10~{\rm kyr}}\right),\label{energy}
\ee
where $E_{\max}\sim10^{40}~{\rm erg}$ is the observed maximum isotropic energy. 
Such a burst energy is sustainable by a magnetar 
for a duration of

\be
\tau\lesssim17~{\rm yr}~\eta_{-3} B_{15}^2,
\ee 
even if the efficiency factor is higher, e.g. $\eta\sim10^{-3}$. Such a time is much shorter than the typical age of a magnetar. This suggests that within the magnetar picture, the extremely high FRB event rate only lasts for a relatively short period of time relative to the typical magnetar lifetime of $\sim(1-10)~{\rm kyr}$ \citep[e.g.,][]{Beniamini19}. Active FRB repeaters in the universe therefore may originate from very young and highly magnetized magnetars \citep[e.g.][]{Metzger17,Lu20}. 

We consider that FRBs are triggered by crust fracturing of a magnetar during its active phase. Since the outer crust has a much lower density than that in the inner crust, relatively frequent starquakes can occur in this region due to the shear stress by magnetic fields. In the following discussion, we focus on the region of outer crust.
In the outer crust with a thickness of $L\sim10^4~{\rm cm}$, the typical density is  $\rho\sim 10^{-3}\rho_{\rm nuc}$, where $\rho_{\rm nuc}=2.8\times10^{14}~{\rm g~cm^{-3}}$ is the nuclear density, and $Y_e\sim0.3$ is the electron fraction \citep{Cumming04}. 
When the shear stress of the magnetic field $B\delta B/4\pi$ reaches a critical value, the crust would crack, leading to starquakes and further triggering bursting radiation (FRB or XRB) by releasing the magnetic energy. The critical condition for magnetic shear stress reaching the threshold is given by

\be
\frac{B\delta B}{4\pi}\simeq\sigma_{\rm cr}\equiv\theta_{\max}\mu,\label{dB}
\ee
where $\theta_{\max}$ is the maximum yield strain of the crust, the shear modulus of the lattice is \citep[e.g.,][]{Piro05}

\be
\mu\simeq1.0\times10^{28}~{\rm erg~cm^{-3}}\rho_{\rm nuc,-3}^{4/3},
\ee
and $\rho_{\rm nuc,-3}=\rho/10^{-3}\rho_{\rm nuc}$, which is normalized to the typical value of an NS outer crust \citep{Douchin01}. Here, we take $Z=40$ as the number of protons per ion, $A=130$ as the number of nucleons per ion, and $X_n=0$ as the fraction of neutrons outside nuclei \citep{Douchin01}. 
We define the bent angle of magnetic field as $\theta_B=\delta B/B$. According to Eq.(\ref{dB}), the critical value of the bent angle is 

\be
\theta_{B,c}=\left(\frac{4\pi\mu}{B^2}\right)\theta_{\max}=10^{-3}~{\rm rad}~\mu_{28}B_{15}^{-2}\theta_{\max,-2}.\label{bent}
\ee
During the evolution of the magnetic fields, once the magnetic bent angle exceeds $\theta_{B,c}$, the outer crust would be fractured. In the following discussion, we will analyze the magnetic field evolution in the outer crust.

The magnetic field evolution in plasma can be generally written as \citep[e.g.,][]{Goldreich92,Thompson96,Cumming04}

\be
\frac{\partial\bm{B}}{\partial t}=\nabla\times(\bm{v}\times\bm{B})-\nabla\times(\eta\nabla\times\bm{B})-\nabla\times\left(\frac{\bm{J}\times\bm{B}}{n_ee}\right),\nonumber\\\label{Bevolution}
\ee
where $\eta=c^2/4\pi\sigma$ is the magnetic diffusivity, $\sigma$ the electrical conductivity, and $\bm{J}=(c/4\pi)(\nabla\times\bm{B})$ the current density. In the outer crust, the advection and Ohmic terms (first two terms on the right hand side of Eq.(\ref{Bevolution})) could be neglected since the plasma is almost at rest and that the electrical conductivity is extremely large. Thus, the magnetic field evolution is dominated by the Hall drift term (the third term in the right hand side of Eq.(\ref{Bevolution})) in short term. 
The electron fluid drifts with respect to ions, and the magnetic field is carried by the electron fluid. Although Hall drift is non-dissipative, it can change the geometric configuration of the magnetic field. The magnetic field evolution term due to Hall drift is given by

\be
\left. \frac{\partial\bm{B}}{\partial t} \right |_{\rm Hall} =-\nabla\times\left(\frac{\bm{J}\times\bm{B}}{en_e}\right),
\ee
and the Hall timescale is given by

\be
\tau_{\rm Hall}=\frac{en_eL}{J}=\frac{4\pi en_eL^2}{cB},
\ee
where $n_e=Y_e \rho/m_p$ is the electron density in the outer crust, and $L\simeq P/\rho g$ is the typical length scale over which $B$ and $n_e$ vary, which is just of the order of magnitude of the outer crust thickness of $\sim 10^4~{\rm cm}$. Taking the typical values of the outer crust, one can estimate the Hall timescale as

\be
\tau_{\rm Hall}\simeq11~{\rm yr}~Y_{e,-1}\rho_{\rm nu,-3}L_4^2B_{15}^{-1},
\ee
during which the magnetic field changes by $\delta B\sim B$.
Since the critical bent angle $\theta_{B,c}$ of magnetic fields is small (Eq.(\ref{bent})), the typical timescale triggering crustal cracking would be much shorter than the Hall drift timescale, $\Delta t_{\rm cra}\sim\theta_{B,c}\tau_{\rm Hall} \ll \tau_{\rm Hall}$. The burst rate is then estimated as 

\be
\mathcal{R}=\Delta t_{\rm cra}^{-1}\simeq0.26~{\rm day}^{-1}~Y_{e,-1}^{-1}\rho_{\rm nu,-3}^{-7/3}L_4^{-2}B_{15}^{3}\theta_{\max,-2}^{-1}.\nonumber\\\label{crack}
\ee
One can see that the burst rate satisfies $\mathcal{R}\propto B^3$, suggesting that more frequent repeating FRBs originate from magnetars with stronger magnetic fields within this picture.

In long term, the magnetic field in the crust would decay via Ohmic dissipation. Two mechanisms dominate this process: 1. the nonlinear Hall term gives rise to a turbulent cascade to smaller scales; 2. the Ohmic dissipation rate (the second term of Eq.(\ref{Bevolution})) is enhanced by Hall cascades.
The long-term magnetic field evolution by Hall cascades can be approximated by 
$dB/dt\simeq-AB^2$ with $A=10^{-18}~{\rm G^{-1}yr^{-1}}$ \citep{Colpi00} so that the magnetic decay may be quantified as

\be
B=\frac{B_0}{1+AB_0t},\label{decay}
\ee
where $B_0$ is the initial magnetic field in the crust. The typical timescale for magnetic field decay is

\be
\tau_B=\frac{1}{AB_0}=100~{\rm yr}~B_{0,16}^{-1}
\ee
If $t\gg\tau_B$, the field decay satisfies $B\simeq B_0(t/\tau_B)^{-1}$. According to Eq.(\ref{crack}) and Eq.(\ref{decay}), for a given magnetar, the event rate satisfies

\be
\mathcal{R}\propto
\left\{
\begin{aligned}
&t^0,~&{\rm for}~t\ll\tau_B\\
&t^{-3},~&{\rm for}~t\gg\tau_B. 
\end{aligned}
\right.
\ee
Therefore, active repeating FRBs (i.e., FRBs with high bursting rates) are proposed to originate from young magnetars with strong magnetic fields when $t < \tau_B$. Galactic magnetars typically have $t \gtrsim \tau_B$, so that their burst rate drops significantly, consistent with their low FRB rate as indicated by the only detection from SGR J1935+2154 in several years since CHIME became online.
Since the number of young magnetars in the universe is relatively small, we expect that active repeating FRBs are rarer compared with apparent
one-off FRBs, which are also repeaters but with a much lower repetition rate within this picture. 
Such a result may also offer an interpretation to the periodic activities of FRB 180916B (\citet{CHIME20b,Pastor-Marazuela20} and the possible periodic activity of XRBs of SGR 1806-20 \citep{Zhang21b}), since the internal strong magnetic field of a young magnetar can deform the star, leading to possible free precession with a long period \citep{Zanazzi20,Levin20,Li21} (see, however, \cite{Ioka20b} and \cite{Lyutikov20} for an alternative interpretation of the periodicity due to the binary motion of the FRB-generating magnetar).
 
\begin{figure}[]
    \centering
	\includegraphics[width = 1\linewidth]{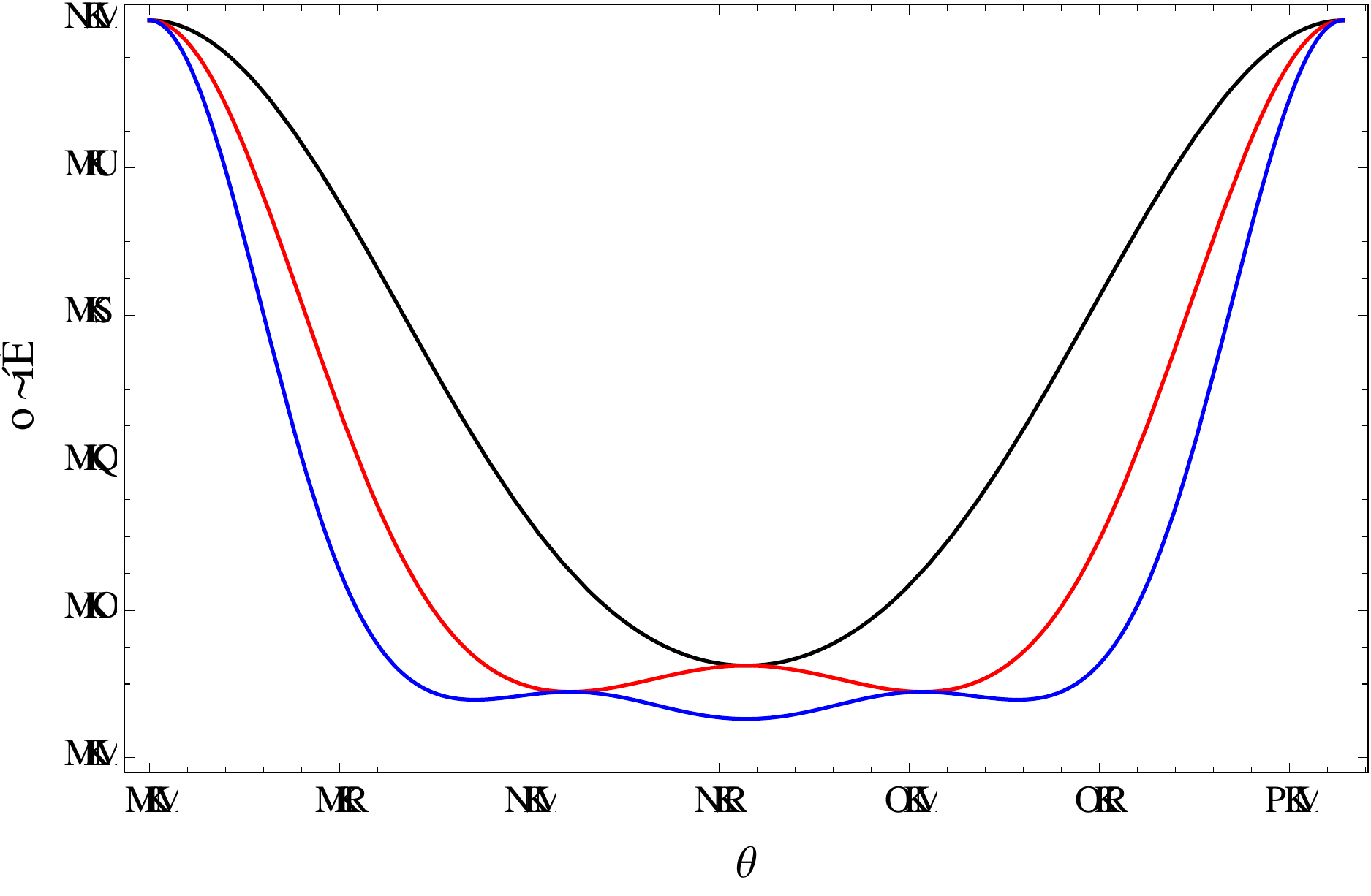}
    \caption{The fracturing rate as a function of  polar angle $\theta$ for a multipole field with $m=0$ and $l={\rm arbitrary}$. The black, red, blue lines corresponds to $l=1,2,3$, respectively. The unit of rate is arbitrary.}\label{fig2}
\end{figure}

\begin{figure}[]
    \centering
	\includegraphics[width = 1\linewidth , trim = 150 150 50 100, clip]{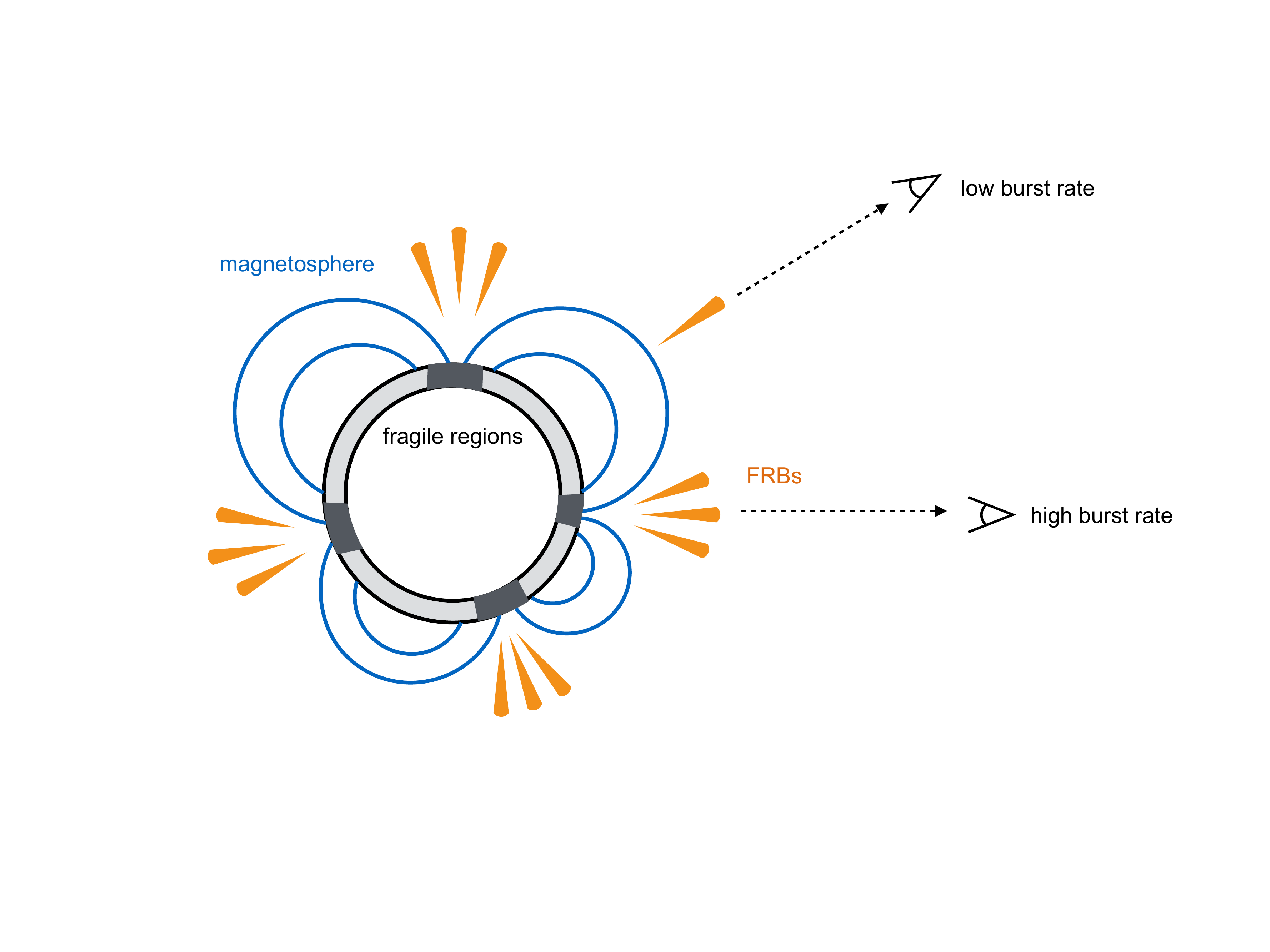}
    \caption{A schematic view of a magnetar with a multipolar magnetic field configuration near surface. At polar regions, FRBs are more easily triggered by crust fracturing. When the fragile regions sweep across the line of sight during magnetar rotation, the source would appear to have a high repetition rate. When the fragile regions are always outside the line of sight, the burst rate would be  much lower.}\label{fig3}
\end{figure}

On the other hand, the repeating behavior may also depend on the line of sight if the FRB emission is beamed, which is naturally expected in a magnetar magnetosphere. Since the polar regions have much stronger fields, crust fracturing may more easily occur in these regions.
The near-surface magnetic field configuration is not well known, and it is possible that there exist multipolar fields near the magnetar crust.
We generally delineate the multipolar magnetic field configuration by spherical harmonics $(m,l)$.
For a multipole field, the magnetic field vector $\bm{B}^{lm}(r,\theta,\varphi)=[B^{lm}_r(r,\theta,\varphi),B^{lm}_\theta(r,\theta,\varphi),B^{lm}_\varphi(r,\theta,\varphi)]$ satisfies \citep[e.g.,][]{Jackson75}

\be
B^{lm}_r(r,\theta,\varphi)&=&4\pi\frac{l+1}{2l+1}\frac{q_{lm}}{r^{l+2}}Y_{lm}(\theta,\varphi), \nonumber\\
B^{lm}_\theta(r,\theta,\varphi)&=&-\frac{4\pi}{2l+1}\frac{q_{lm}}{r^{l+2}}\frac{\partial}{\partial\theta}Y_{lm}(\theta,\varphi), \nonumber\\
B^{lm}_\varphi(r,\theta,\varphi)&=&-\frac{4\pi}{2l+1}\frac{q_{lm}}{r^{l+2}}\frac{im}{\sin\theta} Y_{lm}(\theta,\varphi).
\ee
For simplicity, we assume an axisymmetric multipolar magnetic field with $m=0$ and $l={\rm arbitrary~number}$, leading to $B^{lm}_\varphi=0$, and $Y_{l0}=\sqrt{(2l+1)/4\pi}P_l(\cos\theta)$, where $P_l(x)$ is Legendre polynomial. At the crust, the strength of magnetic field at the polar angle $\theta$ is 

\be
B=\sqrt{\frac{4\pi}{2l+1}}\frac{q_{l0}}{R^{l+2}}\sqrt{(l+1)^2P_l(\cos\theta)^2+P'_l(\cos\theta)^2}\nonumber\\
\ee
where the prime symbol represents the derivative with respect to $\theta$. According to Eq.(\ref{crack}), the fracturing rate satisfies $\mathcal{R}\propto B^3$. Assuming  that the scale $L$ is independent of $\theta$ in the crust, one may estimate the fracturing rate as 

\be
\mathcal{R}(\theta)\propto\left[(l+1)^2P_l(\cos\theta)^2+P'_l(\cos\theta)^2\right]^{3/2}.
\ee
As shown in Figure \ref{fig2}, the fracturing rate is relatively larger at the main poles of a multipolar field, and the $\theta$-range of a crust fracturing region is smaller for a multipole field with a larger $l$. 
We call the region with a large fracturing rate as a ``fragile region''. 
A real multipole field may be composed of sub components with different $(m,l)$ so that there are multiple fragile regions on the magnetar crust, as shown in the schematic picture in Figure \ref{fig3}.
If the fragile regions sweep across the line of sight as the magnetar rotate, one would detect FRBs with a high repetition rate. Recently, \citet{Li21b} found that the bursts of FRB 121102 appears to have two components in the energy distribution function. According to this picture, this might correspond to two fragile regions sweeping the line of sight. Due to the different magnetic field strengths at these fragile regions, they show different typical energies and burst rate distributions. 
On the other hand, if all fragile regions are always outside the line of sight, the burst rate would be much lower. In the latter case, if some bursts from fragile regions are slightly off-axis, their observed fluxes would be suppressed by the beaming structure or the Doppler effect \citep{Zhang21}, which might correspond to the case of SGR J1935+2154 with a lower luminosity and a lower burst rate (based on the observation time of FRB 200428). 

\section{Coherent plasma radiation of FRBs}\label{coherent} 

The extremely high brightness temperature of FRBs, typically $T_B\sim10^{35}~{\rm K}$, implies that the radiation mechanism of FRBs must be coherent. Since FRBs have some properties similar to radio emission of normal pulsars, several pulsar-like coherent radiation mechanisms have been proposed (\cite{Zhang20} and references therein). The connections between FRBs and radio pulsars are manifested by the following facts: 
\begin{itemize}
\item FRBs and radio pulsars are the most coherent sources in all kinds of astrophysical phenomena. 
In addition to normal radio pulses, some pulsars show giant pulses with the brightness temperature comparable to FRBs \citep[e.g.,][]{Hankins07}, although the luminosity of giant pulses is much smaller than that of FRBs.
\item Both periodic radio pulses and FRB 200428 were emitted from Galactic magnetar SGR J1935+2154 \citep{Zhu20}. Meanwhile, there have been some bright radio bursts with fluxes between those of pulsed emission and of FRB emission detected from this magnetar \citep{Zhang20b,Kirsten20,Good20}. 
\end{itemize}

Based on the observed flux and frequency of an FRB, one can estimate the electric field strength $E_w$ of the electromagnetic wave at a distance $r$ from the center of the magnetar as 

\be
E_w&\simeq&\left(\frac{4\pi\nu F_\nu}{c}\right)^{1/2}\frac{d}{r}\nonumber\\
&\simeq&6\times10^{8}~{\rm V~cm^{-1}}\nu_{\rm GHz}^{1/2}F_{\nu,{\rm Jy}}^{1/2}d_{\rm Gpc} r_7^{-1},\label{Efrb}
\ee
where $d$ is the distance between the FRB source and the observer. 
We consider the generation of FRB radiation in the magnetosphere of a magnetar. According to \citet{Kumar20} and \citet{Lu20}, Alfv\'en waves are induced by the crust cracking in the polar region and propagate along the magnetic field lines. Since the plasma density decreases with the radius, a strong parallel electric field may be induced as Alfv\'en waves reach a critical radius of $r_{\rm cr}\sim 10R$ where charge starvation occurs \citep{Kumar20,Lu20}, as shown in panel (a) of Figure \ref{fig1}.
Different from \citet{Kumar20} and \citet{Lu20} who invoked the accelerating electric field to continuously accelerate bunches to emit coherent radiation, in the following we consider a pair cascade in the charge starved region, which further generates coherent plasma radiation.

When the crust cracks, an magnetic field disturbance is produced by the wiggling of the magnetic field foot points by the crust fracturing motion, i.e.

\be
\delta B&\sim&\frac{v}{c}B_p\sim\frac{1}{c}\left(\frac{\mu}{\rho}\right)^{1/2}B_p\nonumber\\
&\simeq&6\times10^{12}~{\rm G}~B_{p,15}\rho_{\rm nu,-3}^{1/6},
\ee
where $v$ is comparable to the shear wave velocity in the outer crust, 

\be
v\sim v_\mu=\left(\frac{\mu}{\rho}\right)^{1/2}=6\times10^{-3}c~\rho_{\rm nu,-3}^{1/6}.\label{shear}
\ee
The Alfv\'en wave packet travels along magnetic field lines and its amplitude $\delta B$ decreases with distance, $\delta B(r)\propto r^{-3/2}$. At the charge starved region $r_{\rm cr}$, the parallel electric field is \citep{Kumar20}

\be
E_{\rm gap}&=&\frac{k_{\perp}}{k_{\parallel}}\left(\frac{r_{\rm rc}}{R}\right)^{-3/2}\delta B\nonumber\\
&=&2\times10^{10}~{\rm V~cm^{-1}}~\eta_{k,-1}B_{p,15}\rho_{\rm nu,-3}^{1/6}r_{\rm cr,7}^{-3/2}\nonumber\\\label{Egap2}
\ee
where $k_\perp$ and $k_\parallel$ are the components of the wavevector of Alfv\'en waves perpendicular and parallel to the background magnetic field, respectively, and 

\be
\eta_{k}\equiv \frac{k_\perp}{k_\parallel}\sim0.1
\ee
at $r_{\rm cr}\sim10 R$ \citep{Kumar20}.

In the charge starved region near at $r_{\rm cr}\sim 10 R$, due to the existence of $E_{\rm gap}$, electrons and positrons would be accelerated and emit gamma-ray photons due to curvature radiation or inverse Compton scattering off the thermal photons \citep{Zhang00}. The gamma-ray photons produce new pairs via the QED processes\footnote{Photon splitting may suppress pair production when $B\gtrsim B_Q$ \citep{Baring01}. However, if only the E$\rightarrow$OO mode is operating, then there is no suppression of pair yields with photon splitting alone \citep{Shabad86,Harding06}.} $\gamma+B\rightarrow e^++e^-$ \citep{Daugherty96} or $\gamma+\gamma \rightarrow e^++e^-$ with thermal photons near the magnetar surface \citep{Zhang01}\footnote{There should be $\gamma$-ray emission accompanied from this processs. However, the luminosity of this emission component is comparable to that of the FRB, which is much dimmer than the other XRB-emission mechanism as discussed in Sect.\ref{Xray}.}. 
Then the cascade process is interrupted when the fresh pairs generated by the cascade process is able to screen the accelerating electric field $E_{\rm gap}$. 
When $E_{\rm gap}$ is screened, the plasma still carries a finite current that reverses the initial electric field. This process could be repeated, driving a inductive oscillations as proposed in the pulsar scenario\footnote{In the active magnetar scenario, the oscillation of Alfv\'en wave in the magnetosphere might also cause an oscillating electric field at the same time.} \citep{Levinson05,Timokhin13,Cruz21}. 

\citet{Philippov20} performed a simulation and claimed that coherent electromagnetic radiation can be directly generated in non-stationary pair plasma discharges. The mechanism was applied to interpret pulsar radio emission. 
Consider that the accelerating electric field in the gap region is $E_{\rm gap}$, then the amplitude of electrostatic wave is $E_x\sim \xi_LE_{\rm gap}$ with the parameter $\xi_L\sim(0.1-1)$ for large-amplitude oscillations (e.g., \cite{Levinson05,Timokhin13,Cruz21}, also see the simulation of \citet{Philippov20}). 
For this coherent radiation mechanism, the most important condition to generate electromagnetic waves is nonuniform pair creation across magnetic field lines. Define $\Theta$ as the angle between the normal to plasma injection front and the background magnetic field. The amplitude of electromagnetic wave is then about $\sim E_x\sin\Theta\sim \xi_LE_{\rm gap}\sin\Theta$. For coherent radiation from pulsar polar gap, nonuniform pair creation across magnetic field lines is due to the different degrees of acceleration at different positions in the polar cap, finally leading to $\Theta\sim0.1$ \citep{Philippov20}. Therefore, the amplitude of electromagnetic waves is about $\sim 0.01E_{\rm gap}$ as shown by the simulation results presented in Figure 1 of \citet{Philippov20}.

\begin{figure}[]
    \centering
	\includegraphics[width = 1.1\linewidth , trim = 80 150 10 150, clip]{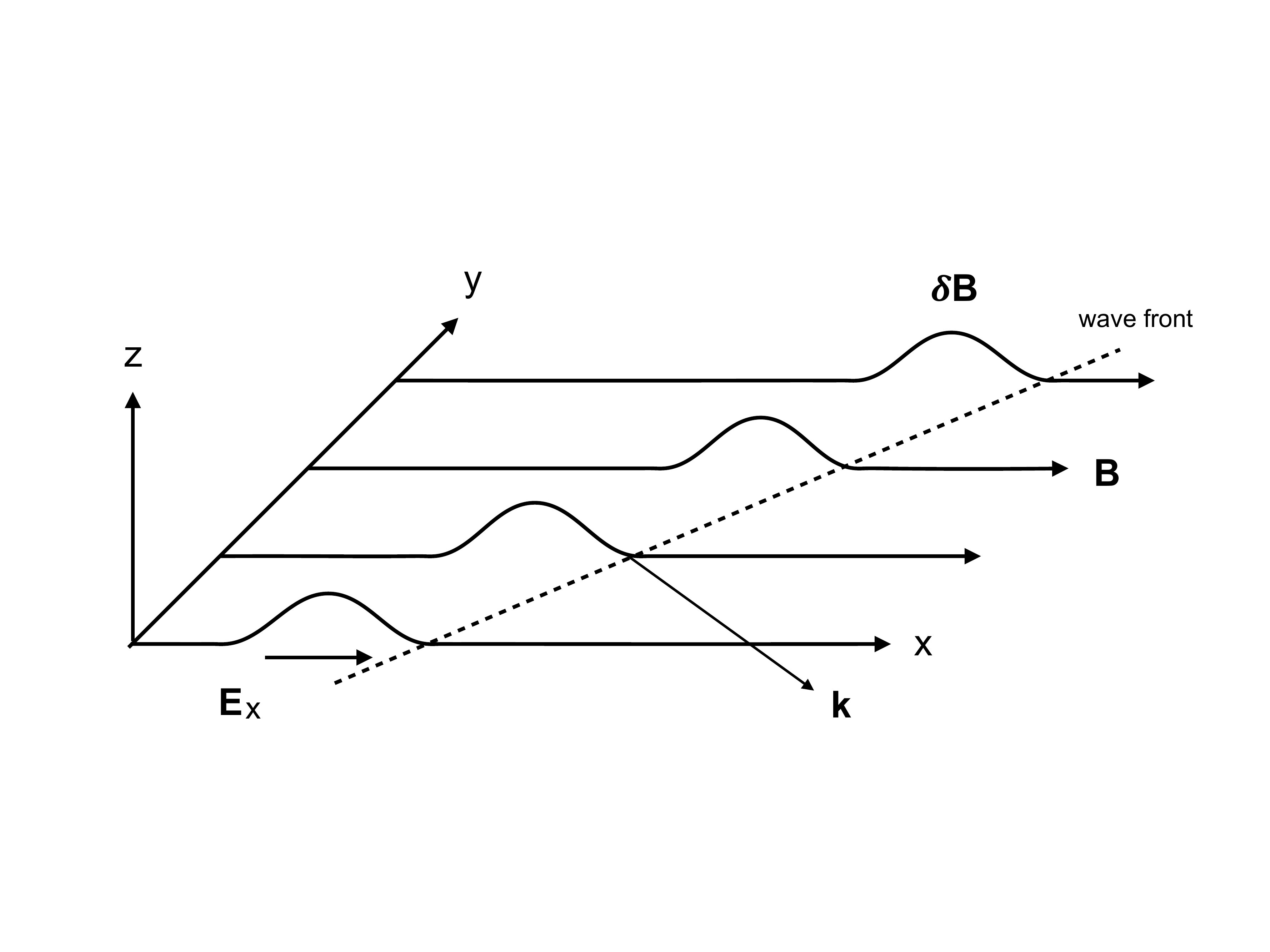}
    \caption{The transversal displacements of plasma and magnetic field in the Alfv\'en wave. The dashed line denote the wave front of Alfv\'en wave.}\label{fig4}
\end{figure}

In the following discussion, we will give a semi-quantitative analysis for this mechanism and apply it to FRBs.
First, we consider that the discharge occurs in the background magnetic field along $x$-axis, similar to \cite{Philippov20}. Particles move along the field line and the electric current is $j_x$. The magnetic field is assumed to be uniform in the $z$-direction, leading to $\partial_z=0$ for all quantities, as shown in Figure \ref{fig4}. For an E-mode wave, since its field does not couple with the plasma, it cannot be directly emitted in the discharging process. For an O-mode wave, the Maxwell's equations give 

\be
\partial_yB_z&=&\frac{4\pi}{c}j_x+\frac{1}{c}\partial_tE_x,\nonumber\\
-\partial_xB_z&=&\frac{1}{c}\partial_tE_y,\nonumber\\
\partial_xE_y-\partial_yE_x&=&-\frac{1}{c}\partial_tB_z,\label{Maxwell}
\ee
which suggests that O-mode waves are coupled with the plasma because of the $j_x$ term. Meanwhile, transverse modes can be only excited when the system is non-uniform in the $y$-direction. 
Therefore, the non-uniform discharge in the $y$-direction is the necessary condition to generate electromagnetic waves. Based on Eq.(\ref{Maxwell}), one obtains

\be
\partial_t\left(\frac{1}{c^2}\partial_tE_x+\frac{4\pi}{c^2}j_x\right)+\partial_y\left(\partial_xE_y-\partial_yE_x\right)=0.\label{efield}
\ee
Injection of particles due to pair formation gives rise to the electric current $j_x$ driven by the accelerating electric field in the gap, and $j_x$ induces a fluctuating electric field $E_x$. 
In the charge starved region, the parallel electric field $E_{x}$ still dominates plasma processes, leading to

\be
\partial_tE_x+4\pi j_x\simeq0.\label{lfield}
\ee
For an electron in a one-dimensional relativistic plasma, its equation of motion is 

\be
\frac{d(\gamma m_ev)}{dt}\sim(\gamma^3 m_e)\frac{v}{\tau_p}\simeq eE_x,\label{mass}
\ee
where $\gamma^3m_e$ corresponds to the ``longitudinal mass'' of the electron in one-dimensional plasma due to $d(\gamma v)=\gamma^3dv$. Using Eq.(\ref{lfield}), Eq.(\ref{mass}) and the definition of the current $j_x=n_eev$, the plasma oscillation frequency is estimated as

\be
\omega_p=\frac{1}{\tau_p}=\left(\frac{4\pi e^2 n_e}{\gamma^3m_e}\right)^{1/2}
=5.6~{\rm GHz}~n_{e,16}\gamma_2^{-3/2}\label{pfrequency}
\ee
where $n_e$ is the electron density in the emission region\footnote{This density of $n_e\sim10^{16}~{\rm cm^{-3}}$ is consistent with the density of pair flow entailed by XRB at $r\sim10R$, see \citet{Ioka20} that focuses on the discussion of chocked scenario of FRB in active magnetosphere. Notice that here $\gamma$ is the local Lorentz factor of electron by accelerated electric field.}. 
If $\partial_y=0$, according to Eq.(\ref{efield}) and Eq.(\ref{lfield}), there would be no electromagnetic wave generated. However, for $\partial_y\neq0$ (implying an inevitable nonuniformity of pair formation across the magnetic field lines), the non-uniformity of $E_x$ across magnetic field lines would give rise to the fluctuating magnetic field $B_z$, and further induce fluctuating perpendicular electric field $E_y$. The electric field $E_y$ directly reflects the amplitude of the transverse wave, and the typical frequency of the coherent radiation of the transverse wave is $\nu\sim\omega_p/2\pi$ \citep{Philippov20}.
According to Eq.(\ref{efield}) and Eq.(\ref{lfield}), for $\partial_y\neq0$, one obtains the relation between $E_x$ and $E_y$, i.e.

\be
\partial_xE_y-\partial_yE_x\simeq0.\label{key}
\ee

For the Alfv\'en waves in the magnetosphere, the components of wavevector are $k_\perp$ and $k_\parallel$, respectively. Thus, the angle $\Theta$ between the background magnetic field direction and the normal direction to the plasma injection front satisfies $\sin\Theta=k_\perp/k_\parallel=\eta_k$. Based on Eq.(\ref{key}), the amplitude of transverse wave is

\be
E_w\sim E_y\simeq\frac{k_\perp}{k_\parallel}\xi_L E_{\rm gap}\label{Ew}
\ee
Since the parallel electric field with large-oscillation amplitude is dominant in the $x$-direction, one has $E_x\sim \xi_L E_{\rm gap}$ with parameter $\xi_L\sim (0.1-1)$ (see the simulation of \citep{Philippov20}). This suggests that an ordinary mode is excited in the charge starved region, and the amplitude of the electromagnetic wave is $E_w\sim E_y$.
For an FRB with typical parameters, according to Eq.(\ref{Efrb}), Eq.(\ref{Egap2}), and Eq.(\ref{Ew}), $\xi_L\sim0.3$ is required, which is consistent with the simulation result.
Therefore, the predicted isotropic energy of FRBs is estimated as 

\be
E_{\rm FRB,iso}&=&f_b \left(\frac{cE_w^2}{4\pi}\right)\Delta r^2\Delta t\nonumber\\
&=&9\times10^{37}~{\rm erg}~f_{b,1}\eta_{k,-1}^4\xi_{L,-1}^2\beta_{-2}^2 B_{p,15}^2r_{\rm cr,7}^{-3}\Delta r_7^3,\label{Enfrb}\nonumber\\
\ee
where $f_b=4\pi/\Delta\Omega$ which is normalized tp 10, $\Delta\Omega$ is the radiation beaming angle, $\beta=v/c$, $v$ is the shear wave velocity given by Eq.(\ref{shear}), $\Delta t\sim\Delta r/c$ is the duration, $\Delta r$ is the scale of the FRB emission region. The energy given by Eq.(\ref{Enfrb}) is consistent with the energy distribution of extragalactic FRBs \citep{Luo20,Lu20}.

On the other hand, as proposed by \citet{Thompson01}, the high-energy radiation, i.e. X-ray or gamma-ray emission, is driven by internal magnetic stresses in the crust. The creation of fireballs by Alfv\'en waves outside the star involves the rapid relaxation of the external magnetic stresses. 
Assuming that the fireball scale $\Delta R$ (notice that it is different from the FRB emission scale $\Delta r$) is comparable to the neutron star radius, i.e. $\Delta R \sim R$, one may estimate the energy powering the XRB as

\be
E_{\rm XRB}\sim\frac{\delta B^2}{8\pi}\Delta R^3=4.0\times10^{42}~{\rm erg}~\delta B_{13}^2\Delta R_6^3.\nonumber\\
\ee
Taking $\delta B\sim\beta B$, then the energy ratio between the FRB and the XRB is 

\be
\eta=\frac{E_{\rm FRB, iso}}{E_{\rm XRB}}\sim2\times10^{-5}\eta_{k,-1}^4\xi_{L,-1}^2r_{\rm cr,7}^{-3}\Delta r_7^3\Delta R_6^{-3} \nonumber\\
\ee
for the above other typical parameters.
The ratio depends on the properties of Alfv\'en waves, pair cascade processes, and size of the trapped fireball.
In conclusion, magnetars with a stronger field tends to generate FRBs with a larger burst energy and a higher event rate, which implies that extragalactic active repeating FRB sources are from younger magnetars with stronger fields than Galactic magnetars.

\section{X-ray emission from an FRB source}\label{Xray}

\subsection{X-ray emission properties associated with FRB 200428}

FRB 200428 with two peaks separated by $\sim$ 30 ms \citep{Bochenek20,CHIME20} was detected from SGR J1935+2154 during its active phase in association with a hard X-ray burst \citep{Li20,Mereghetti20,Ridnaia20,Tavani20}. 
The FRB-associated XRB showed two hard X-ray peaks whose arrival times are consistent with the two FRB peaks after de-dispersion \citep{Li20}.
The delay time of the X-ray peak from the FRB pulse was $\sim6.5$ ms, but the onset of the X-ray burst started $\sim 30$ ms before the FRB \citep{Li20,Mereghetti20}.
Meanwhile, the associated XRB appears to have some different properties from normal XRBs, including a sharp peak with millisecond duration and a harder energy cutoff \citep{Li20,Mereghetti20,Younes20}. 

Based on the estimated distance of SGR J1935+2154 \citep{Pavlovic13,Kothes18,Zhou20,Zhong20}, the isotropic energy of FRB 200428 is $E_{\rm FRB, iso}\sim10^{35}~{\rm erg}$, and the energy of the associated X-ray burst is $E_{\rm XRB}\sim10^{40}~{\rm erg}$ (XRBs from magnetars are widely considered to be nearly isotropic).
Thus, the ratio between the radio and X-ray fluxes is $\eta= E_{\rm FRB, iso}/E_{\rm XRB}\sim10^{-5}$.

The FRB-associated XRB is considered to be generated from trapped fireballs and/or relativistic outflows from an active magnetar magnetosphere \citep[e.g.,][]{Ioka20,Yamasaki20b}.
Since the onset of the XRB started 30 ms before the FRB, whether an FRB can break out from the magnetosphere becomes an important issue. \cite{Ioka20} argued that an electron pair outflow from a trapped-expanding fireball would be entailed by the associated XRB, which would further pollute the magnetosphere before the FRB emission. 
Although the pair flow is opaque to induced Compton scattering of FRB photons due to the extremely high brightness temperature of the FRB, the FRB can break out of the pair outflow by a radiation force if the FRB emission radius is larger than a few tens of magnetar radii. 
The critical condition is given by $L_{\rm FRB}>\tau_{\rm T}L_{\rm XRB}$ \citep{Ioka20}, where $L_{\rm FRB}$ and $L_{\rm XRB}$ are the luminosities of the FRB and the XRB, respectively, $\tau_{\rm T}$ is the Thomson optical depth in the FRB emission region.
Based on the above critical condition, a breakout radius of a few tens of $R$ is obtained for FRB 200428 \citep{Ioka20}, which is consistent with the radius of charge starved region of FRB as discussed in Section \ref{coherent}. 
On the other hand, for a given FRB emission radius, if the luminosity ratio between an FRB and its associated XRB becomes smaller, the FRB would be chocked in the magnetosphere, which might explain the low association rate by deep monitoring of FAST observations \citep{Lin20}.

In the following discussion, we focus on the properties of the sharp-peak component of the associated XRB of FRB 200428, including the sharp peaks and a harder spectral cutoff. We take a typical value $B_p=10^{14}~{\rm G}$, which is more consistent with that of SGR J1935+2154. 

\subsection{Resonant Compton scattering process during X-ray emission}

Resonant Compton scattering has been studied to explain the spectra of magnetar emission \citep[e.g.,][]{Thompson02,Baring07,Beloborodov07,Fernandez07,Baring20,Yamasaki20}. In the following, we focus on the resonant Compton scattering process of the escaping photons from the bottom of a trapped fireball.
Under the strong magnetic field of a magnetar, X-ray photons have two polarization modes: The O-mode has the electric vector in the plane of the wavevector and the background magnetic field, and its cross section is about $\sigma_{\rm O}\sim\sigma_{\rm T}$; and the E-mode has an electric vector perpendicular to this plane and its cross section is B-dependent,
$\sigma_{\rm E}\simeq\left(\omega/\omega_B\right)^2\sigma_{\rm T}$,
where $\omega_B=eB/m_ec$ is the cyclotron frequency. The stronger the magnetic field, the more transparent the magnetosphere to the E-mode. 
Therefore, when a trapped fireball forms, the E-mode X-ray photons would escape from the bottom of trapped fireball near the magnetar surface because 
$\sigma_{\rm E}\propto B^{-2}\propto r^{6}$, as shown in panel (b) of Figure \ref{fig1}.

Once the E-mode X-ray photons escape from the base of a trapped fireball, they will enter a region between two adjacent trapped fireballs, and further be resonantly scattered by the electrons in this region.
The frequency-dependent cross section of resonant Compton scattering is 

\be
\sigma_{\rm res}(\omega')\simeq2\pi^2 r_ec\delta(\omega'-\omega_B)=\sigma_{\rm res,0}\omega_B\delta(\omega'-\omega_B),\nonumber\\\label{res}
\ee
where $\omega'$ is the electromagnetic wave frequency in the electron rest frame, $r_e=e^2/m_ec^2$ is the classical electron radius, and $\sigma_{\rm res,0}=2\pi^2e/B$.
At the resonant frequency, the cross section of resonant Compton scattering is much larger than that of Compton scattering in the magnetosphere,
$\sigma_{\rm res,0}/\sigma_{\rm T}\sim (1/\alpha)\left(B/B_Q\right)^{-1}$,
where $\alpha\simeq1/137$ is the fine structure constant. We consider that the strength of the dipole magnetic field depends on radius $r$, i.e. $B\simeq B_p(r/R)^{-3}$. Thus the critical radius is
$r_{\rm res}\sim\left(\alpha B_p/B_Q\right)^{1/3}R$.
The resonant Compton scattering would be dominant at $r\gg r_{\rm res}$. For a given incident photon energy,  
the typical resonant Lorentz factor for resonant Compton scattering is \citep[e.g.,][]{Beloborodov07}

\be
\gamma_{\rm res}&=&\frac{1}{1-\cos\theta_{kB}}\frac{B}{B_Q}\left(\frac{\epsilon}{m_ec^2}\right)^{-1}\nonumber\\
&=&512f(\theta_{kB})(B/B_Q)\epsilon_{\rm keV}^{-1},\label{gamres}
\ee
where $\epsilon=\hbar\omega=1~{\rm keV}\epsilon_{\rm keV}$ is the energy of the incident photon, $\theta_{kB}$ is the angle between the incident photon and the magnetic field, $f(\theta_{kB})=1/(1-\cos\theta_{kB})$. 
The radiative force exerted on an electron near the magnetar surface is estimated by

\be
F_r&\simeq&\int\sigma_{\rm res}(\omega)\frac{L_\omega}{4\pi r^2c}d\omega=\frac{\pi e L_X}{2cB_pR^2}\nonumber\\
&=&2.5\times10^{-5}~{\rm dyne}~L_{X,41}B_{p,14}^{-1},
\ee
where $\sigma_{\rm res}(\omega)$ is the cross section of resonant Compton scattering given by Eq.(\ref{res}), $\omega'=\gamma(1-\cos\theta_{kB})\omega$, and $L_\omega\sim L_X/\omega$.
After the FRB is emitted (with a delay time of 6 ms), the radiative force would change the electron momentum distribution. 
The relaxation timescale for electrons to acquire the equilibrium momentum distribution can be estimated by $\gamma m_ec/\tau_{\rm rel}\sim F_r$, which gives

\be
\tau_{\rm rel}&\sim&\frac{\gamma m_ec}{F_r}\simeq\frac{2\gamma m_ec^2B_pR^2}{\pi eL_X}\nonumber\\
&\simeq&1.1\times10^{-12}~{\rm s}~\gamma L_{X,41}^{-1}B_{p,14}.
\ee
This means that the momentum distribution of electrons changes in an extremely short period of time. \citet{Yamasaki20} proposed that resonant Compton scattering of the balanced relativistic electrons would change the X-ray spectrum, leading to a higher cut-off energy. In the electron rest frame,
the incident photon angle with respect to the local magnetic field is 

\be
\cos\theta_{kB}'=\frac{\cos\theta_{kB}-\beta}{1-\beta\cos\theta_{kB}}.
\ee
Consider an electron that initially moves in parallel with the magnetic field with $\beta\sim0$. It would be accelerated by the radiation force. 
Once the electron has a relativistic velocity $\beta\sim1$, it would be decelerated by the radiation force, because the incident photon is from $\cos\theta_{kB}'\sim-1$ in the electron rest frame. 
Therefore, the electron under radiative force would be immediately accelerated or decelerated to reach an equilibrium state. In the equilibrium state, the radiation force is directed perpendicularly to the magnetic field so that $\theta_{kB}'=\pi/2$. 
The balancing velocity of the electron under radiative force is

\be
\beta_{\rm res}=\cos\theta_{kB};~~~~~\gamma_{\rm res}=\frac{1}{\sin\theta_{kB}}.\label{vres}
\ee
Using Eq.(\ref{gamres}) and Eq.(\ref{vres}). The resonant Compton energy of incident photon is 

\be
\epsilon=\frac{\sin\theta_{kB}}{1-\cos\theta_{kB}}\frac{B}{B_Q}m_ec^2.
\ee
For a magnetar with $B_p\sim 10B_Q$, at a distance $r\sim 5R$, the energy of the incident photon is required to be $\epsilon\simeq40~{\rm keV}g(\theta_{kB})$, where $g(\theta_{kB})=\sin\theta_{kB}/(1-\cos\theta_{kB})$. 
The energy of the scattered photon is about

\be
\epsilon_s\sim\gamma_{\rm res}^2 \epsilon
\ee
for $\theta_{kB}\gtrsim1$. 
Since $\gamma_{\rm res}\sim$ a few, according to Eq.(\ref{vres}), the cutoff energy of X-ray emission would be increased by several times compared with those without resonant Compton scattering, which might explain the high-energy cutoff of the associated XRB of FRB 200428.
A detailed discussion about the spectral modification could be found in \citep{Yamasaki20}, which has been applied to model the spectra of giant flares and even intermediate flares of magnetars. 

Another important observation is that the light curve of the FRB-associated XRB showed sharp peaks aligned with the FRB pulses \citep{Li20}. We propose that the sharp-peak component is from the region between two adjacent trapped fireballs, as shown in panel (b) of Figure \ref{fig1}. 
For E-mode X-ray photons, the radiative flux across the trapped fireball surface is concentrated close to the bottom of the trapped fireball near the magnetar surface, and further escapes along more extended magnetic field lines in the region between two adjacent trapped fireballs. A large flux is emitted in this region and the radiation beam is confined by the trapped fireballs, leading to the sharp-peak component in light curve. The effect of the confined beam is discussed as follow.

As discussed in Section \ref{fracture}, during the magnetar active phase, the magnetic field near the magnetar surface may have a general multipolar configuration described by $(m,l)$. For simplicity, we still assume an axisymmetric multipolar magnetic field with $m=0$ and $l={\rm arbitrary}$.
The first-order approximation for the field line equation is \citep[e.g.,][]{Asseo02}

\be
r\sim R_{\max}\sin^{2/l}\theta
\ee
for $\theta\lesssim1$, where $\theta$ is the angle from the multipolar pole. Then the opening angle of a field line at one pole is estimated as

\be
\theta_{\rm open}\sim\left(\frac{R}{R_{\max}}\right)^{l/2}.\label{theopen}
\ee
For example, for a closed field line with $R_{\max}\sim10^4R$, a dipole field ($l=1$) has an opening angle of $\theta_{\rm open}\sim0.01~{\rm rad}$, but a tripole field ($l=2$) has $\theta\sim10^{-4}~{\rm rad}$. Thus, multipolar fields have much narrower opening angles than dipolar fields. 

When the X-ray photons enter the region between two adjacent trapped fireballs, they will be confined in a beaming angle due to the large optical depth $\tau\gg1$ at the top region of the trapped fireball around it, as shown in panel (b) of Figure \ref{fig1}. For two adjacent trapped fireballs with a size of $\Delta R$, one has $R_{\max}\sim R+\Delta R$. According to Eq.(\ref{theopen}), the scale of the region between two adjacent trapped fireballs is estimated by 

\be
l_{\rm beam}=R\theta_{\rm open}\sim R\left(\frac{R}{R+\Delta R}\right)^{l/2}.
\ee
Consider that X-ray photons escape from the base of a trapped fireball. Since $\sigma_{\rm E}\propto r^6$, the upper regions of the adjacent trapped fireballs are opaque to X-rays. Thus, the beaming angle is about

\be
\theta_{\rm beam}\sim\frac{l_{\rm beam}}{\Delta R}\simeq\frac{R}{\Delta R}\left(\frac{R}{R+\Delta R}\right)^{l/2}.
\ee
We assume that the energy of a trapped fireball is $\Delta E\sim E_{\rm XRB}\simeq10^{40}~{\rm erg}$ and the temperature is about $kT\sim10~{\rm keV}$, then the size of the trapped fireball is about $\Delta R\sim R$.
For a trapped fireball with $\Delta R\sim R$ and $l=4$, the beaming angle of X-ray emission is $\theta_{\rm beam}=0.25~{\rm rad}$. 
Therefore, the radiation from this region will be confined in a narrow beam, especially for multipolar fields near the magnetar surface.

\subsection{Temperature evolution of persistent X-ray emission}\label{temperature}

In addition to XRBs, SGR J1935+2154 also showed persistent emission with a double-peaked pulse profile with the rotation period of the magnetar during the magnetar active phase. The arrival time of FRB 200428 aligned in phase with the brightest peak of the pulse profile \citep{Younes20}. The blackbody temperature of the X-ray persistent emission decreases rapidly in the early stage of the outburst, while the size of the emitting area remains the same with a radius of a few km. In this section, we propose that the X-ray persistent emission is from a hot spot on the magnetar surface and calculate its temperature evolution. See panel (b) of Figure \ref{fig1}.

In the crust of a magnetar, the Coulomb lattice of heavy nuclei coexists in $\beta$-equilibrium with a degenerate electron gas.
The Fermi energy of degenerate electrons in a strong field is \citep[e.g.,][]{Ruderman71,Flowers77,Lai01}

\be
\epsilon_{F,e}=\frac{2\pi^4\hbar^4c^2Y_e^2\rho^2}{e^2m_p^2m_eB^2}=180~{\rm GeV}~Y_{e,-1}^{2}\rho_{\rm nuc,-3}^{2}B_{15}^{-2},\nonumber\\
\ee
where $B$ is the magnetic field in magnetar crust.
Since the fraction of the effective thermal electrons is only $\sim kT/\epsilon_{F,e}$ for a Fermi electron gas, the internal energy by relativistic electrons is

\be
U_e&\simeq&\frac{1}{2}N_ekT\left(\frac{kT}{\epsilon_{F,e}}\right)\nonumber\\
&=&7.3\times10^{30}~{\rm erg}~Y_{e,-1}^{-1}\rho_{\rm nuc,-3}^{-1}T_{\rm keV}^2L_4\Delta R_{\rm hs,5}^2B_{15}^2\nonumber\\
\ee
where $N_e\simeq n_eL\Delta R_{\rm hs}^2$ is the total electron number in the hot spot with size  $\Delta R_{\rm hs}$ and depth $L$, $\rho\sim10^{-3}\rho_{\rm nuc}$ is the density of the outer crust, $T=1~{\rm keV} \ T_{\rm keV}/k$ is the temperature of the hot spot, and the factor of $1/2$ is due to the one-dimensional freedom of electrons in strong fields. On the other hand, the ions in the outer crust are nondegenerate, and each ion has thermal energy of $(3/2)k T$. Thus, the internal energy stored by ions in the outer crust is

\be 
U_i&=&\frac{3}{2}N_ikT\nonumber\\
&=&3.1\times10^{38}~{\rm erg}~\rho_{\rm nuc,-3}T_{\rm keV}L_4\Delta R_{\rm hs,5}^2,
\ee
where $N_i\simeq (\rho/Am_p)L\Delta R_{\rm hs}^2$ is the total ion number with $A=130$. 

Therefore, the thermal energy of the hot spot is dominated by ions, $U\simeq U_i$, and such an energy could be produced by $\sim1\%$ of an XRB. The temperature evolution is given by
$dU_i/dt=-\Delta R_{\rm hs}^2\sigma T^4$.
The typical cooling time is defined as

\be
t_{\rm cool}=\frac{U_i}{\Delta R_{\rm hs}^2\sigma T^4}=0.3~{\rm day}~\rho_{\rm nuc,-3}^{2/3}T_{\rm keV}^{-3}L_4,
\ee
which is consistent with the rapid temperature evolution of the X-ray persistent emission, as shown in Figure 8 of \cite{Younes20}. Meanwhile, the cooling time from  temperature $T_0$ to $T$ is given by

\be
\Delta t=t-t_0=\frac{k\rho L}{2Am_p\sigma}\left(\frac{1}{T^3}-\frac{1}{T_0^3}\right),
\ee 
where $T_0$ is the initial temperature at time $t_0$.

\section{Conclusion and Discussion}\label{summary}

Prompted by the observed association between FRB 200428 and its X-ray counterpart from Galactic magnetar SGR J1935+2154 \citep{Bochenek20,CHIME20,Li20,Mereghetti20,Ridnaia20,Tavani20}, we propose that FRBs and XRBs are triggered by crust fracturing of magnetars. Due to Hall drift, the geometric configuration of the magnetic fields would evolve in the crust. When the shear stress reaches a critical value depending on the shear modulus, the outer crust would crack and further trigger magnetosphere activities. We find that the event rate of crust fracturing depends on the magnetic field strength, i.e. $\mathcal{R}\propto B^3$. 
Since the magnetic field in the crust decays by Ohmic dissipation and Hall cascades in long term, our results suggest that the active repeating FRB sources are younger magnetars with stronger magnetic fields. 
On the other hand, the observed repeating behavior also depends on the line of sight if the FRB emission is beamed.  
If the fragile regions sweep across the line of sight as the magnetar rotates, one would see FRBs with a high repetition rate. However, if the fragile regions are always the line of sight, the burst rate would be much lower. 

According to \citet{Kumar20} and \citet{Lu20}, a charge starved region could be generated by Alfv\'en waves  triggered by crust fracturing after propagating to a distance of $\sim 10R$. Meanwhile, the association between an FRB and an XRB also implies that the emission region is higher than $\sim 10R$ in order for an FRB to break out from the magnetosphere \citep{Ioka20}.
Non-stationary pair plasma discharges would be produced in this charge starved region, similar to inner gap sparking theorized in radio pulsars \citep{Ruderman75} but with a higher emission region at $\sim10R$. Inspired by the simulations of \citet{Philippov20}, we consider coherent plasma radiation as the radiation mechanism of FRBs due to the similarity of the coherent radio emission between FRBs and normal radio pulsars. Near the charge starved region, nonuniform pair production across magnetic field lines would lead to electromagnetic waves whose amplitude is proportional to the strength of the induced electric field. This electromagnetic wave would be strong enough to be an FRB. According to this mechanism, 
magnetars with stronger fields tend to generate FRBs with larger burst energies and a higher event rate. 

On the other hand, the sharp-peak component of the FRB-associated XRB is proposed to be generated in the region between adjacent trapped fireballs. The E-mode X-ray photons escape from the base of the trapped fireballs and get resonantly scattered to higher energies, leading to the observed high cutoff energy of the FRB-associated XRB \citep[e.g.,][]{Li20,Younes21}. Meanwhile, when X-ray photons escape from the bottom of the trapped fireballs  and enter the region between two adjacent trapped fireballs, they are confined in a beaming angle due to the large optical depth $\tau\gg1$ of the trapped fireballs around it. Therefore, a large flux is emitted in this region and the radiation beam is collimated by the trapped fireballs, leading to the sharp-peak component in the light curve. 

After an XRB, a part of energy (about $1\%$) of the magnetosphere activity is injected to the magnetar surface and produces a hot spot. The persistent X-ray emission is suggested to be the thermal emission from this hot spot. Since cooling of the hot spot is dominated by thermal ions on the magnetar surface, the typical cooling timescale is about $\sim0.3~{\rm day}$ for a hot spot with size of a few km and temperature of $\sim1~{\rm keV}$. This cooling timescale is consistent with the observations of SGR J1935+2154 \citep{Younes20}. 

\acknowledgments

We thank the anonymous referee for helpful comments and suggestions.
We also thank Clara Dehman, Jinjun Geng, Ersin Gogus, Qiaochu Li, Zenan Liu, Kohta Murase, Yuanhong Qu, Fayin Wang, Weiyang Wang, Zhaoyang Xia, Shotaro Yamasaki and Jinping Zhu for helpful discussions. Y.P.Y is supported by National Natural Science Foundation of China grant No. 12003028.


\begin{thebibliography}{}
\expandafter\ifx\csname natexlab\endcsname\relax\def\natexlab#1{#1}\fi

\bibitem[{{Asseo} \& {Khechinashvili}(2002)}]{Asseo02}
{Asseo}, E., \& {Khechinashvili}, D. 2002, \mnras, 334, 743

\bibitem[{{Baring} \& {Harding}(2001)}]{Baring01}
{Baring}, M.~G., \& {Harding}, A.~K. 2001, \apj, 547, 929

\bibitem[{{Baring} \& {Harding}(2007)}]{Baring07}
---. 2007, \apss, 308, 109

\bibitem[{{Baring} {et~al.}(2020){Baring}, {Wadiasingh}, {Gonthier}, {Harding},
  \& {Hu}}]{Baring20}
{Baring}, M.~G., {Wadiasingh}, Z., {Gonthier}, P.~L., {Harding}, A.~K., \&
  {Hu}, K. 2020, arXiv e-prints, arXiv:2012.10815

\bibitem[{{Beloborodov}(2017)}]{Beloborodov17}
{Beloborodov}, A.~M. 2017, \apjl, 843, L26

\bibitem[{{Beloborodov} \& {Thompson}(2007)}]{Beloborodov07}
{Beloborodov}, A.~M., \& {Thompson}, C. 2007, \apj, 657, 967

\bibitem[{{Beniamini} {et~al.}(2019){Beniamini}, {Hotokezaka}, {van der Horst},
  \& {Kouveliotou}}]{Beniamini19}
{Beniamini}, P., {Hotokezaka}, K., {van der Horst}, A., \& {Kouveliotou}, C.
  2019, \mnras, 487, 1426

\bibitem[{{Bochenek} {et~al.}(2020){Bochenek}, {Ravi}, {Belov}, {Hallinan},
  {Kocz}, {Kulkarni}, \& {McKenna}}]{Bochenek20}
{Bochenek}, C.~D., {Ravi}, V., {Belov}, K.~V., {et~al.} 2020, \nat, 587, 59

\bibitem[{{Chatterjee} {et~al.}(2017){Chatterjee}, {Law}, {Wharton},
  {Burke-Spolaor}, {Hessels}, {Bower}, {Cordes}, {Tendulkar}, {Bassa},
  {Demorest}, {Butler}, {Seymour}, {Scholz}, {Abruzzo}, {Bogdanov}, {Kaspi},
  {Keimpema}, {Lazio}, {Marcote}, {McLaughlin}, {Paragi}, {Ransom}, {Rupen},
  {Spitler}, \& {van Langevelde}}]{Chatterjee17}
{Chatterjee}, S., {Law}, C.~J., {Wharton}, R.~S., {et~al.} 2017, \nat, 541, 58

\bibitem[{{CHIME/FRB Collaboration} {et~al.}(2020{\natexlab{a}}){CHIME/FRB
  Collaboration}, {Andersen}, {Bandura}, {Bhardwaj}, {Bij}, {Boyce}, {Boyle},
  {Brar}, {Cassanelli}, {Chawla}, {Chen}, {Cliche}, {Cook}, {Cubranic},
  {Curtin}, {Denman}, {Dobbs}, {Dong}, {Fandino}, {Fonseca}, {Gaensler},
  {Giri}, {Good}, {Halpern}, {Hill}, {Hinshaw}, {H{\"o}fer}, {Josephy},
  {Kania}, {Kaspi}, {Landecker}, {Leung}, {Li}, {Lin}, {Masui}, {McKinven},
  {Mena-Parra}, {Merryfield}, {Meyers}, {Michilli}, {Milutinovic},
  {Mirhosseini}, {M{\"u}nchmeyer}, {Naidu}, {Newburgh}, {Ng}, {Patel}, {Pen},
  {Pinsonneault-Marotte}, {Pleunis}, {Quine}, {Rafiei-Ravandi}, {Rahman},
  {Ransom}, {Renard}, {Sanghavi}, {Scholz}, {Shaw}, {Shin}, {Siegel}, {Singh},
  {Smegal}, {Smith}, {Stairs}, {Tan}, {Tendulkar}, {Tretyakov}, {Vanderlinde},
  {Wang}, {Wulf}, \& {Zwaniga}}]{CHIME20}
{CHIME/FRB Collaboration}, {Andersen}, B.~C., {Bandura}, K.~M., {et~al.}
  2020{\natexlab{a}}, \nat, 587, 54

\bibitem[{{CHIME/FRB Collaboration} {et~al.}(2020{\natexlab{b}}){CHIME/FRB
  Collaboration}, {Amiri}, {Andersen}, {Bandura}, {Bhardwaj}, {Boyle}, {Brar},
  {Chawla}, {Chen}, {Cliche}, {Cubranic}, {Deng}, {Denman}, {Dobbs}, {Dong},
  {Fandino}, {Fonseca}, {Gaensler}, {Giri}, {Good}, {Halpern}, {Hessels},
  {Hill}, {H{\"o}fer}, {Josephy}, {Kania}, {Karuppusamy}, {Kaspi}, {Keimpema},
  {Kirsten}, {Landecker}, {Lang}, {Leung}, {Li}, {Lin}, {Marcote}, {Masui},
  {McKinven}, {Mena-Parra}, {Merryfield}, {Michilli}, {Milutinovic},
  {Mirhosseini}, {Naidu}, {Newburgh}, {Ng}, {Nimmo}, {Paragi}, {Patel}, {Pen},
  {Pinsonneault-Marotte}, {Pleunis}, {Rafiei-Ravandi}, {Rahman}, {Ransom},
  {Renard}, {Sanghavi}, {Scholz}, {Shaw}, {Shin}, {Siegel}, {Singh}, {Smegal},
  {Smith}, {Stairs}, {Tendulkar}, {Tretyakov}, {Vanderlinde}, {Wang}, {Wang},
  {Wulf}, {Yadav}, \& {Zwaniga}}]{CHIME20b}
{CHIME/FRB Collaboration}, {Amiri}, M., {Andersen}, B.~C., {et~al.}
  2020{\natexlab{b}}, \nat, 582, 351
  
\bibitem[The CHIME/FRB Collaboration et al.(2021)]{CHIME21} The CHIME/FRB Collaboration, :, Amiri, M., et al.\ 2021, arXiv:2106.04352


\bibitem[{{Colpi} {et~al.}(2000){Colpi}, {Geppert}, \& {Page}}]{Colpi00}
{Colpi}, M., {Geppert}, U., \& {Page}, D. 2000, \apjl, 529, L29

\bibitem[{{Cordes} \& {Chatterjee}(2019)}]{Cordes19}
{Cordes}, J.~M., \& {Chatterjee}, S. 2019, \araa, 57, 417

\bibitem[{{Cruz} {et~al.}(2021){Cruz}, {Grismayer}, \& {Silva}}]{Cruz21}
{Cruz}, F., {Grismayer}, T., \& {Silva}, L.~O. 2021, \apj, 908, 149

\bibitem[{{Cumming} {et~al.}(2004){Cumming}, {Arras}, \& {Zweibel}}]{Cumming04}
{Cumming}, A., {Arras}, P., \& {Zweibel}, E. 2004, \apj, 609, 999

\bibitem[{{Daugherty} \& {Harding}(1996)}]{Daugherty96}
{Daugherty}, J.~K., \& {Harding}, A.~K. 1996, \apj, 458, 278

\bibitem[Dehman et al.(2020)]{Dehman20} Dehman, C., Vigan{\`o}, D., Rea, N., et al.\ 2020, \apjl, 902, L32. doi:10.3847/2041-8213/abbda9

\bibitem[{{Douchin} \& {Haensel}(2001)}]{Douchin01}
{Douchin}, F., \& {Haensel}, P. 2001, \aap, 380, 151

\bibitem[{{Fern{\'a}ndez} \& {Thompson}(2007)}]{Fernandez07}
{Fern{\'a}ndez}, R., \& {Thompson}, C. 2007, \apj, 660, 615

\bibitem[{{Flowers} \& {Ruderman}(1977)}]{Flowers77}
{Flowers}, E., \& {Ruderman}, M.~A. 1977, \apj, 215, 302

\bibitem[{{Geng} {et~al.}(2021){Geng}, {Li}, \& {Huang}}]{Geng21}
{Geng}, J.-J., {Li}, B., \& {Huang}, Y.-F. 2021, arXiv e-prints,
  arXiv:2103.04165

\bibitem[{{Goldreich} \& {Reisenegger}(1992)}]{Goldreich92}
{Goldreich}, P., \& {Reisenegger}, A. 1992, \apj, 395, 250

\bibitem[{{Good} \& {CHIME/FRB Collaboration}(2020)}]{Good20}
{Good}, D., \& {CHIME/FRB Collaboration}. 2020, The Astronomer's Telegram,
  14074, 1

\bibitem[{{Gourdji} {et~al.}(2019){Gourdji}, {Michilli}, {Spitler}, {Hessels},
  {Seymour}, {Cordes}, \& {Chatterjee}}]{Gourdji19}
{Gourdji}, K., {Michilli}, D., {Spitler}, L.~G., {et~al.} 2019, \apjl, 877, L19

\bibitem[{{Hankins} \& {Eilek}(2007)}]{Hankins07}
{Hankins}, T.~H., \& {Eilek}, J.~A. 2007, \apj, 670, 693

\bibitem[{{Harding} \& {Lai}(2006)}]{Harding06}
{Harding}, A.~K., \& {Lai}, D. 2006, Reports on Progress in Physics, 69, 2631

\bibitem[{{Ioka}(2020)}]{Ioka20}
{Ioka}, K. 2020, \apjl, 904, L15

\bibitem[{{Ioka} \& {Zhang}(2020)}]{Ioka20b}
{Ioka}, K., \& {Zhang}, B. 2020, \apjl, 893, L26

\bibitem[{{Jackson}(1975)}]{Jackson75}
{Jackson}, J.~D. 1975, {Classical electrodynamics}

\bibitem[{{Katz}(2016)}]{Katz16}
{Katz}, J.~I. 2016, \apj, 826, 226

\bibitem[{{Kirsten} {et~al.}(2020){Kirsten}, {Snelders}, {Jenkins}, {Nimmo},
  {van den Eijnden}, {Hessels}, {Gawro{\'n}ski}, \& {Yang}}]{Kirsten20}
{Kirsten}, F., {Snelders}, M.~P., {Jenkins}, M., {et~al.} 2020, Nature
  Astronomy, arXiv:2007.05101

\bibitem[{{Kothes} {et~al.}(2018){Kothes}, {Sun}, {Gaensler}, \&
  {Reich}}]{Kothes18}
{Kothes}, R., {Sun}, X., {Gaensler}, B., \& {Reich}, W. 2018, \apj, 852, 54

\bibitem[{{Kumar} \& {Bo{\v{s}}njak}(2020)}]{Kumar20}
{Kumar}, P., \& {Bo{\v{s}}njak}, {\v{Z}}. 2020, \mnras, 494, 2385

\bibitem[{{Kumar} {et~al.}(2017){Kumar}, {Lu}, \& {Bhattacharya}}]{Kumar17}
{Kumar}, P., {Lu}, W., \& {Bhattacharya}, M. 2017, \mnras, 468, 2726

\bibitem[{{Lai}(2001)}]{Lai01}
{Lai}, D. 2001, Reviews of Modern Physics, 73, 629

\bibitem[{{Law} {et~al.}(2017){Law}, {Abruzzo}, {Bassa}, {Bower},
  {Burke-Spolaor}, {Butler}, {Cantwell}, {Carey}, {Chatterjee}, {Cordes},
  {Demorest}, {Dowell}, {Fender}, {Gourdji}, {Grainge}, {Hessels}, {Hickish},
  {Kaspi}, {Lazio}, {McLaughlin}, {Michilli}, {Mooley}, {Perrott}, {Ransom},
  {Razavi-Ghods}, {Rupen}, {Scaife}, {Scott}, {Scholz}, {Seymour}, {Spitler},
  {Stovall}, {Tendulkar}, {Titterington}, {Wharton}, \& {Williams}}]{Law17}
{Law}, C.~J., {Abruzzo}, M.~W., {Bassa}, C.~G., {et~al.} 2017, \apj, 850, 76

\bibitem[Levin \& Lyutikov(2012)]{Levin12} Levin, Y. \& Lyutikov, M.\ 2012, \mnras, 427, 1574. doi:10.1111/j.1365-2966.2012.22016.x

\bibitem[{{Levin} {et~al.}(2020){Levin}, {Beloborodov}, \&
  {Bransgrove}}]{Levin20}
{Levin}, Y., {Beloborodov}, A.~M., \& {Bransgrove}, A. 2020, \apjl, 895, L30

\bibitem[{{Levinson} {et~al.}(2005){Levinson}, {Melrose}, {Judge}, \&
  {Luo}}]{Levinson05}
{Levinson}, A., {Melrose}, D., {Judge}, A., \& {Luo}, Q. 2005, \apj, 631, 456

\bibitem[{{Li} {et~al.}(2020){Li}, {Lin}, {Xiong}, {Ge}, {Li}, {Li}, {Lu},
  {Zhang}, {Tuo}, {Nang}, {Zhang}, {Xiao}, {Chen}, {Song}, {Xu}, {Liu}, {Jia},
  {Cao}, {Qu}, {Zhang}, {Gu}, {Liao}, {Zhao}, {Tan}, {Nie}, {Zhao}, {Zheng},
  {Zheng}, {Luo}, {Cai}, {Li}, {Xue}, {Bu}, {Chang}, {Chen}, {Chen}, {Chen},
  {Chen}, {Cui}, {Cui}, {Deng}, {Dong}, {Du}, {Fu}, {Gao}, {Gao}, {Gao}, {Gu},
  {Guan}, {Guo}, {Han}, {Huang}, {Huo}, {Jiang}, {Jiang}, {Jin}, {Jin}, {Kong},
  {Li}, {Li}, {Li}, {Li}, {Li}, {Li}, {Li}, {Liang}, {Liu}, {Liu}, {Liu},
  {Liu}, {Liu}, {Lu}, {Lu}, {Luo}, {Ma}, {Meng}, {Ou}, {Sai}, {Shang}, {Song},
  {Sun}, {Tao}, {Wang}, {Wang}, {Wang}, {Wang}, {Wang}, {Wen}, {Wu}, {Wu},
  {Wu}, {Xiao}, {Xu}, {Yang}, {Yang}, {Yang}, {Yang}, {Yi}, {Yin}, {You},
  {Zhang}, {Zhang}, {Zhang}, {Zhang}, {Zhang}, {Zhang}, {Zhang}, {Zhang},
  {Zhang}, {Zhang}, {Zhang}, {Zhang}, {Zhang}, {Zhang}, {Zhang}, {Zhang},
  {Zhou}, {Zhou}, {Zhu}, {Zhu}, \& {Zhuang}}]{Li20}
{Li}, C.~K., {Lin}, L., {Xiong}, S.~L., {et~al.} 2020, arXiv e-prints,
  arXiv:2005.11071

\bibitem[{{Li} {et al.}(2021a)}]{Li21c} Li, Q.-C., Yang, Y.-P., Wang, F.Y., Xu, K, Shao, Y., Liu, Z.-N., Dai, Z.-G., 2021, submit

\bibitem[{{Li} {et~al.}(2021b){Li}, {Wang}, {Zhu}, {Zhang}, {Zhang}, \&
  {Duan}}]{Li21b}
{Li}, D., {Wang}, P., {Zhu}, W.~W., {et~al.} 2021, Nature, accepted.

\bibitem[{{Li} \& {Zanazzi}(2021)}]{Li21}
{Li}, D., \& {Zanazzi}, J.~J. 2021, arXiv e-prints, arXiv:2101.05836

\bibitem[{{Lin} {et~al.}(2020){Lin}, {Zhang}, {Wang}, {Gao}, {Guan}, {Han},
  {Jiang}, {Jiang}, {Lee}, {Li}, {Men}, {Miao}, {Niu}, {Niu}, {Sun}, {Wang},
  {Wang}, {Xu}, {Xu}, {Xu}, {Yang}, {Yang}, {Yu}, {Zhang}, {Zhang}, {Zhou},
  {Zhu}, {Castro-Tirado}, {Dai}, {Ge}, {Hu}, {Li}, {Li}, {Li}, {Liang}, {Jia},
  {Querel}, {Shao}, {Wang}, {Wang}, {Wu}, {Xiong}, {Xu}, {Yang}, {Zhang},
  {Zhang}, {Zheng}, \& {Zou}}]{Lin20}
{Lin}, L., {Zhang}, C.~F., {Wang}, P., {et~al.} 2020, \nat, 587, 63

\bibitem[{{Lorimer} {et~al.}(2007){Lorimer}, {Bailes}, {McLaughlin},
  {Narkevic}, \& {Crawford}}]{Lorimer07}
{Lorimer}, D.~R., {Bailes}, M., {McLaughlin}, M.~A., {Narkevic}, D.~J., \&
  {Crawford}, F. 2007, Science, 318, 777

\bibitem[{{Lu} {et~al.}(2020){Lu}, {Kumar}, \& {Zhang}}]{Lu20}
{Lu}, W., {Kumar}, P., \& {Zhang}, B. 2020, \mnras, 498, 1397

\bibitem[{{Luo} {et~al.}(2020){Luo}, {Men}, {Lee}, {Wang}, {Lorimer}, \&
  {Zhang}}]{Luo20}
{Luo}, R., {Men}, Y., {Lee}, K., {et~al.} 2020, \mnras, 494, 665

\bibitem[{{Lyubarsky}(2014)}]{Lyubarsky14}
{Lyubarsky}, Y. 2014, \mnras, 442, L9

\bibitem[{{Lyubarsky}(2021)}]{Lyubarsky21}
---. 2021, arXiv e-prints, arXiv:2103.00470

\bibitem[{{Lyutikov} {et~al.}(2020){Lyutikov}, {Barkov}, \&
  {Giannios}}]{Lyutikov20}
{Lyutikov}, M., {Barkov}, M.~V., \& {Giannios}, D. 2020, \apjl, 893, L39

\bibitem[{{Marcote} {et~al.}(2020){Marcote}, {Nimmo}, {Hessels}, {Tendulkar},
  {Bassa}, {Paragi}, {Keimpema}, {Bhardwaj}, {Karuppusamy}, {Kaspi}, {Law},
  {Michilli}, {Aggarwal}, {Andersen}, {Archibald}, {Bandura}, {Bower}, {Boyle},
  {Brar}, {Burke-Spolaor}, {Butler}, {Cassanelli}, {Chawla}, {Demorest},
  {Dobbs}, {Fonseca}, {Giri}, {Good}, {Gourdji}, {Josephy}, {Kirichenko},
  {Kirsten}, {Landecker}, {Lang}, {Lazio}, {Li}, {Lin}, {Linford}, {Masui},
  {Mena-Parra}, {Naidu}, {Ng}, {Patel}, {Pen}, {Pleunis}, {Rafiei-Ravandi},
  {Rahman}, {Renard}, {Scholz}, {Siegel}, {Smith}, {Stairs}, {Vanderlinde}, \&
  {Zwaniga}}]{Marcote20}
{Marcote}, B., {Nimmo}, K., {Hessels}, J.~W.~T., {et~al.} 2020, \nat, 577, 190

\bibitem[{{Margalit} {et~al.}(2020){Margalit}, {Beniamini}, {Sridhar}, \&
  {Metzger}}]{Margalit20}
{Margalit}, B., {Beniamini}, P., {Sridhar}, N., \& {Metzger}, B.~D. 2020,
  \apjl, 899, L27

\bibitem[{{Mereghetti} {et~al.}(2020){Mereghetti}, {Savchenko}, {Ferrigno},
  {G{\"o}tz}, {Rigoselli}, {Tiengo}, {Bazzano}, {Bozzo}, {Coleiro},
  {Courvoisier}, {Doyle}, {Goldwurm}, {Hanlon}, {Jourdain}, {von Kienlin},
  {Lutovinov}, {Martin-Carrillo}, {Molkov}, {Natalucci}, {Onori}, {Panessa},
  {Rodi}, {Rodriguez}, {S{\'a}nchez-Fern{\'a}ndez}, {Sunyaev}, \&
  {Ubertini}}]{Mereghetti20}
{Mereghetti}, S., {Savchenko}, V., {Ferrigno}, C., {et~al.} 2020, \apjl, 898,
  L29

\bibitem[{{Metzger} {et~al.}(2017){Metzger}, {Berger}, \&
  {Margalit}}]{Metzger17}
{Metzger}, B.~D., {Berger}, E., \& {Margalit}, B. 2017, \apj, 841, 14

\bibitem[{{Michilli} {et~al.}(2018){Michilli}, {Seymour}, {Hessels}, {Spitler},
  {Gajjar}, {Archibald}, {Bower}, {Chatterjee}, {Cordes}, {Gourdji}, {Heald},
  {Kaspi}, {Law}, {Sobey}, {Adams}, {Bassa}, {Bogdanov}, {Brinkman},
  {Demorest}, {Fernandez}, {Hellbourg}, {Lazio}, {Lynch}, {Maddox}, {Marcote},
  {McLaughlin}, {Paragi}, {Ransom}, {Scholz}, {Siemion}, {Tendulkar}, {van
  Rooy}, {Wharton}, \& {Whitlow}}]{Michilli18}
{Michilli}, D., {Seymour}, A., {Hessels}, J.~W.~T., {et~al.} 2018, \nat, 553,
  182

\bibitem[Murase et al.(2016)]{Murase16} Murase, K., Kashiyama, K., \& M{\'e}sz{\'a}ros, P.\ 2016, \mnras, 461, 1498. doi:10.1093/mnras/stw1328


\bibitem[{{Pastor-Marazuela} {et~al.}(2020){Pastor-Marazuela}, {Connor}, {van
  Leeuwen}, {Maan}, {ter Veen}, {Bilous}, {Oostrum}, {Petroff}, {Straal},
  {Vohl}, {Attema}, {Boersma}, {Kooistra}, {van der Schuur}, {Sclocco},
  {Smits}, {Adams}, {Adebahr}, {de Blok}, {Coolen}, {Damstra}, {D{\'e}nes},
  {Hess}, {van der Hulst}, {Hut}, {Ivashina}, {Kutkin}, {Marcel Loose},
  {Lucero}, {Mika}, {Moss}, {Mulder}, {Norden}, {Oosterloo}, {Orr{\'u}},
  {Ruiter}, \& {Wijnholds}}]{Pastor-Marazuela20}
{Pastor-Marazuela}, I., {Connor}, L., {van Leeuwen}, J., {et~al.} 2020, arXiv
  e-prints, arXiv:2012.08348

\bibitem[{{Pavlovi{\'c}} {et~al.}(2013){Pavlovi{\'c}}, {Uro{\v{s}}evi{\'c}},
  {Vukoti{\'c}}, {Arbutina}, \& {G{\"o}ker}}]{Pavlovic13}
{Pavlovi{\'c}}, M.~Z., {Uro{\v{s}}evi{\'c}}, D., {Vukoti{\'c}}, B., {Arbutina},
  B., \& {G{\"o}ker}, {\"U}.~D. 2013, \apjs, 204, 4

\bibitem[{{Petroff} {et~al.}(2019){Petroff}, {Hessels}, \&
  {Lorimer}}]{Petroff19}
{Petroff}, E., {Hessels}, J.~W.~T., \& {Lorimer}, D.~R. 2019, \aapr, 27, 4

\bibitem[{{Petroff} \& {Yaron}(2020)}]{Petroff20}
{Petroff}, E., \& {Yaron}, O. 2020, Transient Name Server AstroNote, 160, 1

\bibitem[{{Petroff} {et~al.}(2016){Petroff}, {Barr}, {Jameson}, {Keane},
  {Bailes}, {Kramer}, {Morello}, {Tabbara}, \& {van Straten}}]{Petroff16}
{Petroff}, E., {Barr}, E.~D., {Jameson}, A., {et~al.} 2016, \pasa, 33, e045

\bibitem[{{Philippov} {et~al.}(2020){Philippov}, {Timokhin}, \&
  {Spitkovsky}}]{Philippov20}
{Philippov}, A., {Timokhin}, A., \& {Spitkovsky}, A. 2020, \prl, 124, 245101

\bibitem[{{Piro}(2005)}]{Piro05}
{Piro}, A.~L. 2005, \apjl, 634, L153

\bibitem[{{Rajwade} {et~al.}(2020){Rajwade}, {Mickaliger}, {Stappers},
  {Morello}, {Agarwal}, {Bassa}, {Breton}, {Caleb}, {Karastergiou}, {Keane}, \&
  {Lorimer}}]{Rajwade20}
{Rajwade}, K.~M., {Mickaliger}, M.~B., {Stappers}, B.~W., {et~al.} 2020,
  \mnras, 495, 3551

\bibitem[{{Ridnaia} {et~al.}(2020){Ridnaia}, {Svinkin}, {Frederiks}, {Bykov},
  {Popov}, {Aptekar}, {Golenetskii}, {Lysenko}, {Tsvetkova}, {Ulanov}, \&
  {Cline}}]{Ridnaia20}
{Ridnaia}, A., {Svinkin}, D., {Frederiks}, D., {et~al.} 2020, arXiv e-prints,
  arXiv:2005.11178

\bibitem[{{Ruderman}(1971)}]{Ruderman71}
{Ruderman}, M. 1971, \prl, 27, 1306

\bibitem[{{Ruderman} \& {Sutherland}(1975)}]{Ruderman75}
{Ruderman}, M.~A., \& {Sutherland}, P.~G. 1975, \apj, 196, 51

\bibitem[{{Shabad} \& {Usov}(1986)}]{Shabad86}
{Shabad}, A.~E., \& {Usov}, V.~V. 1986, \apss, 128, 377

\bibitem[{{Spitler} {et~al.}(2016){Spitler}, {Scholz}, {Hessels}, {Bogdanov},
  {Brazier}, {Camilo}, {Chatterjee}, {Cordes}, {Crawford}, {Deneva}, {Ferdman},
  {Freire}, {Kaspi}, {Lazarus}, {Lynch}, {Madsen}, {McLaughlin}, {Patel},
  {Ransom}, {Seymour}, {Stairs}, {Stappers}, {van Leeuwen}, \&
  {Zhu}}]{Spitler16}
{Spitler}, L.~G., {Scholz}, P., {Hessels}, J.~W.~T., {et~al.} 2016, \nat, 531,
  202

\bibitem[{{Sridhar} {et~al.}(2021){Sridhar}, {Metzger}, {Beniamini},
  {Margalit}, {Renzo}, {Sironi}, \& {Kovlakas}}]{Sridhar21}
{Sridhar}, N., {Metzger}, B.~D., {Beniamini}, P., {et~al.} 2021, arXiv
  e-prints, arXiv:2102.06138

\bibitem[{{Tavani} {et~al.}(2020){Tavani}, {Casentini}, {Ursi}, {Verrecchia},
  {Addis}, {Antonelli}, {Argan}, {Barbiellini}, {Baroncelli}, {Bernardi},
  {Bianchi}, {Bulgarelli}, {Caraveo}, {Cardillo}, {Cattaneo}, {Chen}, {Costa},
  {Del Monte}, {Di Cocco}, {Di Persio}, {Donnarumma}, {Evangelista}, {Feroci},
  {Ferrari}, {Fioretti}, {Fuschino}, {Galli}, {Gianotti}, {Giuliani},
  {Labanti}, {Lazzarotto}, {Lipari}, {Longo}, {Lucarelli}, {Magro},
  {Marisaldi}, {Mereghetti}, {Morelli}, {Morselli}, {Naldi}, {Pacciani},
  {Parmiggiani}, {Paoletti}, {Pellizzoni}, {Perri}, {Perotti}, {Piano},
  {Picozza}, {Pilia}, {Pittori}, {Puccetti}, {Pupillo}, {Rapisarda},
  {Rappoldi}, {Rubini}, {Setti}, {Soffitta}, {Trifoglio}, {Trois},
  {Vercellone}, {Vittorini}, {Giommi}, \& {D' Amico}}]{Tavani20}
{Tavani}, M., {Casentini}, C., {Ursi}, A., {et~al.} 2020, arXiv e-prints,
  arXiv:2005.12164

\bibitem[{{Thompson} \& {Duncan}(1995)}]{Thompson95}
{Thompson}, C., \& {Duncan}, R.~C. 1995, \mnras, 275, 255

\bibitem[{{Thompson} \& {Duncan}(1996)}]{Thompson96}
---. 1996, \apj, 473, 322

\bibitem[{{Thompson} \& {Duncan}(2001)}]{Thompson01}
---. 2001, \apj, 561, 980

\bibitem[{{Thompson} {et~al.}(2002){Thompson}, {Lyutikov}, \&
  {Kulkarni}}]{Thompson02}
{Thompson}, C., {Lyutikov}, M., \& {Kulkarni}, S.~R. 2002, \apj, 574, 332

\bibitem[{{Timokhin} \& {Arons}(2013)}]{Timokhin13}
{Timokhin}, A.~N., \& {Arons}, J. 2013, \mnras, 429, 20

\bibitem[{{Wadiasingh} {et~al.}(2020){Wadiasingh}, {Beniamini}, {Timokhin},
  {Baring}, {van der Horst}, {Harding}, \& {Kazanas}}]{Wadiasingh20}
{Wadiasingh}, Z., {Beniamini}, P., {Timokhin}, A., {et~al.} 2020, \apj, 891, 82

\bibitem[{{Wadiasingh} \& {Timokhin}(2019)}]{Wadiasingh19}
{Wadiasingh}, Z., \& {Timokhin}, A. 2019, \apj, 879, 4

\bibitem[Wang \& Zhang(2019)]{Wang19} Wang, F.~Y. \& Zhang, G.~Q.\ 2019, \apj, 882, 108

\bibitem[{{Wang}(2020)}]{Wang20}
{Wang}, J.-S. 2020, \apj, 900, 172

\bibitem[{{Wang} {et~al.}(2018){Wang}, {Luo}, {Yue}, {Chen}, {Lee}, \&
  {Xu}}]{Wang18}
{Wang}, W., {Luo}, R., {Yue}, H., {et~al.} 2018, \apj, 852, 140

\bibitem[{{Wang} {et~al.}(2020){Wang}, {Xu}, \& {Chen}}]{Wang20b}
{Wang}, W.-Y., {Xu}, R., \& {Chen}, X. 2020, \apj, 899, 109

\bibitem[{{Xiao} {et~al.}(2021){Xiao}, {Wang}, \& {Dai}}]{Xiao21}
{Xiao}, D., {Wang}, F., \& {Dai}, Z. 2021, arXiv e-prints, arXiv:2101.04907

\bibitem[Yamasaki et al.(2020a)]{Yamasaki20b} Yamasaki, S., Kashiyama, K., \& Murase, K.\ 2020, arXiv:2008.03634

\bibitem[{{Yamasaki} {et~al.}(2020b){Yamasaki}, {Lyubarsky}, {Granot}, \&
  {G{\"o}{\u{g}}{\"u}{\c{s}}}}]{Yamasaki20}
{Yamasaki}, S., {Lyubarsky}, Y., {Granot}, J., \& {G{\"o}{\u{g}}{\"u}{\c{s}}},
  E. 2020, \mnras, 498, 484

\bibitem[{{Yang} {et~al.}(2021){Yang}, {Zhang}, {Lin}, {Zhang}, {Zhang},
  {Yang}, {Tu}, {Zou}, {Ye}, {Wang}, \& {Dai}}]{Yang21}
{Yang}, Y.-H., {Zhang}, B.-B., {Lin}, L., {et~al.} 2021, \apjl, 906, L12

\bibitem[{{Yang} \& {Zhang}(2018)}]{Yang18}
{Yang}, Y.-P., \& {Zhang}, B. 2018, \apj, 868, 31

\bibitem[{{Yang} {et~al.}(2020){Yang}, {Zhu}, {Zhang}, \& {Wu}}]{Yang20}
{Yang}, Y.-P., {Zhu}, J.-P., {Zhang}, B., \& {Wu}, X.-F. 2020, \apjl, 901, L13

\bibitem[{{Younes} {et~al.}(2020){Younes}, {G{\"u}ver}, {Kouveliotou},
  {Baring}, {Hu}, {Wadiasingh}, {Begi{\c{c}}arslan}, {Enoto},
  {G{\"o}{\u{g}}{\"u}{\c{s}}}, {Lin}, {Harding}, {van der Horst}, {Majid},
  {Guillot}, \& {Malacaria}}]{Younes20}
{Younes}, G., {G{\"u}ver}, T., {Kouveliotou}, C., {et~al.} 2020, \apjl, 904,
  L21

\bibitem[{{Younes} {et~al.}(2021){Younes}, {Baring}, {Kouveliotou},
  {Arzoumanian}, {Enoto}, {Doty}, {Gendreau}, {G{\"o}{\v{g}}{\"u}{\c{s}}},
  {Guillot}, {G{\"u}ver}, {Harding}, {Ho}, {van der Horst}, {Hu}, {Jaisawal},
  {Kaneko}, {LaMarr}, {Lin}, {Majid}, {Okajima}, {Pope}, {Ray}, {Roberts},
  {Saylor}, {Steiner}, \& {Wadiasingh}}]{Younes21}
{Younes}, G., {Baring}, M.~G., {Kouveliotou}, C., {et~al.} 2021, Nature
  Astronomy, arXiv:2006.11358

\bibitem[{{Yu} {et~al.}(2021){Yu}, {Zou}, {Dai}, \& {Yu}}]{Yu21}
{Yu}, Y.-W., {Zou}, Y.-C., {Dai}, Z.-G., \& {Yu}, W.-F. 2021, \mnras, 500, 2704

\bibitem[{{Zanazzi} \& {Lai}(2020)}]{Zanazzi20}
{Zanazzi}, J.~J., \& {Lai}, D. 2020, \apjl, 892, L15

\bibitem[{{Zhang}(2001)}]{Zhang01}
{Zhang}, B. 2001, \apjl, 562, L59

\bibitem[{{Zhang}(2020{\natexlab{a}})}]{Zhang20c}
---. 2020{\natexlab{a}}, \apjl, 890, L24

\bibitem[{{Zhang}(2020{\natexlab{b}})}]{Zhang20}
---. 2020{\natexlab{b}}, \nat, 587, 45

\bibitem[{{Zhang}(2021)}]{Zhang21}
---. 2021, \apjl, 907, L17

\bibitem[{{Zhang} \& {Harding}(2000)}]{Zhang00}
{Zhang}, B., \& {Harding}, A.~K. 2000, \apj, 532, 1150

\bibitem[{{Zhang} {et~al.}(2020){Zhang}, {Jiang}, {Men}, {Wang}, {Xu}, {Xu},
  {Niu}, {Zhou}, {Guan}, {Han}, {Jiang}, {Lee}, {Li}, {Lin}, {Niu}, {Wang},
  {Wang}, {Xu}, {Yu}, {Zhang}, \& {Zhu}}]{Zhang20b}
{Zhang}, C.~F., {Jiang}, J.~C., {Men}, Y.~P., {et~al.} 2020, The Astronomer's
  Telegram, 13699, 1

\bibitem[{{Zhang} {et~al.}(2021){Zhang}, {Tu}, \& {Wang}}]{Zhang21b}
{Zhang}, G.~Q., {Tu}, Z.-L., \& {Wang}, F.~Y. 2021, arXiv e-prints,
  arXiv:2101.07923

\bibitem[{{Zhong} {et~al.}(2020){Zhong}, {Dai}, {Zhang}, \& {Deng}}]{Zhong20}
{Zhong}, S.-Q., {Dai}, Z.-G., {Zhang}, H.-M., \& {Deng}, C.-M. 2020, \apjl,
  898, L5

\bibitem[{{Zhou} {et~al.}(2020){Zhou}, {Zhou}, {Chen}, {Wang}, {Vink}, \&
  {Wang}}]{Zhou20}
{Zhou}, P., {Zhou}, X., {Chen}, Y., {et~al.} 2020, \apj, 905, 99

\bibitem[{{Zhu} {et~al.}(2020){Zhu}, {Wang}, {Zhou}, {Xu}, {Wang}, {Zhang},
  {Feng}, {Han}, {Jiang}, {Lee}, {Di}, {Lin}, {Men}, {Niu}, {Xu}, {Yang}, {Yu},
  {Zhang}, {Zhang}, {Zhang}, {Zhang}, \& {Zhang}}]{Zhu20}
{Zhu}, W., {Wang}, B., {Zhou}, D., {et~al.} 2020, The Astronomer's Telegram,
  14084, 1

\end{thebibliography}
\end{document}